 \documentclass[traditabstract]{aa}

\usepackage{graphicx}
\usepackage{enumerate}
\usepackage{natbib}
\usepackage{txfonts}
\usepackage{verbatim}
\bibpunct{(}{)}{;}{a}{}{,}    

\begin{document}

\title{Three-dimensional non-LTE radiative transfer effects in \ion{Fe}{i} lines}
\subtitle{I. Flux sheet and flux tube geometries}

\author{R.~Holzreuter\inst{1,2}, S.~K.~Solanki\inst{1,3} }

\institute{Max-Planck-Institut f\"ur Sonnensystemforschung, Max-Planck-Str. 2, 37191 Katlenburg-Lindau, Germany \and 
  Institute of Astronomy, ETH Zurich, CH-8093 Zurich, Switzerland \and 
  School of Space Research, Kyung Hee University, Yongin, Gyeonggi 446-701, Korea\\
  \email{holzreuter@astro.phys.ethz.ch}}

\offprints{R.~Holzreuter}

\date{Received $<$date$>$; accepted $<$date$>$}

\abstract {
In network and active region plages, the magnetic field is concentrated into structures often described as flux tubes (FTs) and sheets (FSs). Three-dimensional (3-D) radiative transfer is important for energy transport in these concentrations. It is also expected to be important for diagnostic purposes but has rarely been applied for that purpose.
Using true 3-D, non-local thermodynamic-equilibrium (non-LTE or NLTE) radiative transfer (RT) in FT and FS models, we compute iron line profiles commonly used to diagnose the Sun's magnetic field by using and comparing the results with those obtained from LTE or one-dimensional (1-D) NLTE calculations. 
Employing a multilevel iron atom, we study the influence of several basic parameters such as either FS or FT Wilson depression, wall thickness, radius/width, thermal stratification or magnetic field strength on Stokes $I$ and the polarized Stokes parameters in the thin-tube approximation. 
The use of different levels of approximations of RT (3-D NLTE, 1-D NLTE, LTE) may lead to considerable differences in profile shapes, intensity contrasts, equivalent widths, and the determination of magnetic field strengths. In particular, LTE, which often provides a good approach in planar 1-D atmospheres, is a poor approximation in our flux sheet model for some of the most important diagnostic \ion{Fe}{i} lines ($524.7$ nm, $525.0$ nm, $630.1$ nm, and $630.2$ nm). The observed effects depend on parameters such as the height of line formation, field strength, and internal temperature stratification. Differences between the profile shapes may lead to errors in the determination of magnetic fields on the order of 10\% to 20\%, while errors in the determined temperature can reach $300-400$K. The empirical FT models NET and PLA turn out to minimize the effects of 3-D RT, so that results obtained with these models by applying LTE may also remain valid for 3-D NLTE calculations. Finally, horizontal RT is found to only insignificantly smear out structures such as the optically thick walls of flux tubes and sheets, allowing features as narrow as $10$ km to remain visible.

\keywords{Line: formation -- Magnetic fields -- Polarization -- Radiative transfer --  Sun: atmosphere -- Sun: photosphere}
}

\maketitle

\section{Introduction}\label{sec:fluxsheet_intro}
The magnetic fields in both the solar network and active region plages is concentrated into elements that are generally described by magnetic flux tubes and sheets \citep{stenflo1973,spruit1976,solanki1993}. These magnetic elements form the footpoints of either coronal funnels and loops and chromospheric canopies. As such, they channel the energy transported from the solar interior to the upper solar atmosphere \citep{narainulmschneider1996} and play an important role in modulating the Sun's total and spectral irradiance \citep{domingoetal2009}. 

Magnetic elements are thought to be formed by the magneto-convective processes of flux expulsion and convective collapse. Flux expulsion  separates the magnetic flux from the convection \citep{parker1963a,parker1963b,weiss1966,gallowayweiss1981,hurlburttoomre1988}, while convective collapse brings field strengths up to kG levels \citep[e.g.][]{parker1978, spruit1979, grossmanndoerthetal1998, litesetal2008, danilovicetal2010}. The magnetic elements have a range of sizes down to as small as $100$ km, close to the highest currently achievable resolution \citep{laggetal2010}, and probably lower than that too (Riethm\"uller et al., in prep.).

Radiation plays a dominant role in determining the properties of magnetic elements, both during the formation---i.e. during the convective collapse, where radiation leads to the cooling of the gas in the magnetic features, cf. \citet{venkatakrishnan1986}---and during the normal lifetime of a magnetic element. Since the kG magnetic field suppresses the convection within the magnetic elements, their photospheric layers are heated mainly by radiation. Owing to the evacuation produced by the excess magnetic pressure within the magnetic elements, radiation flows in through the walls and heats the interiors of the magnetic elements \citep{spruit1976}. Multidimensional radiative transfer (RT) plays a central role in setting this temperature and is a standard part of modern MHD simulations of magnetoconvection in the Sun's surface layers \citep{nordlund1982, nordlund1983, nordlund1985rev, deinzeretal1984, deinzeretal1984b, voegleretal2004, voegleretal2005}.

Diagnostic RT, however, is generally carried out in 1-D or at most in so-called 1.5-D, i.e. along many parallel rays passing through a multidimensional model \citep[e.g.][]{buenteetal1993,solanki1989}. This difference in the treatment of the energetics and the diagnostics is a basic inconsistency that is worth it to be investigated. We henceforth also use the term 1-D for 1.5-D RT. Although there have been a number of 3-D RT studies of convective features \citep[][ and references therein]{carlsson2009}, the multi-dimensional RT diagnostics of magnetic elements are to our knowledge restricted to only a few papers \citep{stenholmstenflo1977,stenholmstenflo1978, brulsvdluehe2001}. 

Owing to the very restricted computer capacities at the time, Stenholm and Stenflo in their pioneering papers had to make some compromises, restricting themselves to a strictly cylindrical flux tube and a two-level atom. In the present work, we use a more sophisticated (but still simple) flux-tube model where its expansion with height is computed consistently using the thin-tube (thin flux-sheet) approximation \citep[][ and references therein]{leighton1963,parker1974b,parker1974,spruitroberts1983}, a multi-level atom, and true 3-D NLTE RT. A comparison with 3-D radiation MHD simulations has shown that the thin-tube (or thin flux-sheet) approximation provides a reasonable description of the magnetic flux concentrations in the simulations \citep{yelleschaoucheetal2009}. 

The thin FS geometry was studied using 2-D RT by \citet{brulsvdluehe2001} and part of our work resembles theirs. However, while \citet{brulsvdluehe2001} aimed mainly at establishing that features narrower than the horizontal photon  mean free path can still be resolved, the aim of the present paper is broader. We aim to quantify the influence of multi-dimensional NLTE diagnostic RT relative to LTE and 1-D NLTE RT. To this end, we carry out a parameter study probing the influence of magnetic field strength, flux-tube radius, flux-tube atmospheric model,and UV opacity on the Stokes profiles of four commonly used spectral lines. Although MHD simulations have reached a high level of realism, we prefer to be able to change parameters at will in a simpler model. This allows us to get a clearer understanding of the effects of multi-dimensional RT in comparison with 1-D NLTE and 1-D LTE. 3-D diagnostic RT in hydrodynamic and MHD simulations will be treated in forthcoming papers.

The structure of our work is as follows: In section \ref{sec:fluxsheet_model}, we provide a summary of our model ingredients. In section \ref{sec:fluxsheet_results}, we describe the results of our parameter study in detail. In Section \ref{sec:fluxsheet_discussion}, we discuss the influence of the parameter variations on widely used quantities, such as contrasts and equivalent widths,  focussing on the differences between three calculation methods (LTE, 1-D NLTE, and 3-D NLTE). Our conclusions are presented in section \ref{sec:fluxsheet_conclusions}.

\section{Model ingredients}\label{sec:fluxsheet_model}

\subsection{Radiative transfer calculations}\label{sec:fluxsheet_model_RT}
All our calculations were done with the RT code RH adapted to our needs. This versatile code was developed by \citet{uitenbroek2000} and is based on the iteration scheme of \citet{rybickihummer1991, rybickihummer1992, rybickihummer1994}. It uses an improved version \citep{socasnavarroetal2000} of the short characteristics (SC) method of \citet{olsonkunasz1987} and \citet{kunaszauer1988} for Stokes $I$ and \citet{reesetal1989} for the full Stokes vector. The code is able to do LTE or NLTE calculations in 1-D, 2-D, and 3-D. If not otherwise stated, the calculations presented here, were always done with the 3-D version of the code  \citep{uitenbroek2006}.  In addition to several adaptions for convenience or to enhance the performance, we made a principal change to the formal solver that has a considerable influence on our results. In the appendix, we summarize our improved method for the formal solution used for both Stokes $I$ and full Stokes vector calculations, as well as some pitfalls we encountered with both the old and new methods.

\subsection{Atomic model and iron abundance}\label{sec:fluxsheet_model_atom}
\begin{table}
\caption{$f$-values of our four selected lines}
\begin{center}
\begin{tabular}{ll|ll}
  Line [nm]    & $f$-value  &  Line [nm]  & $f$-value \\
\hline
 524.7           & $5.00\times10^{-06}$  & 630.1           & $4.26\times10^{-02}$   \\
 525.0           & $3.00\times10^{-05}$  & 630.2           & $2.47\times10^{-02}$   \\
\hline
\end{tabular}
\end{center}
\label{tab:f-values}
\end{table}

We used an \ion{Fe}{i} atom  including 22 \ion{Fe}{i} levels and the ground state of \ion{Fe}{ii}, which allowed us to perform a realistic NLTE calculation of selected multiplets containing transitions of interest. The model, kindly provided by J.\ H.\ M.\ J.\ Bruls, employs energy level data compiled by \citet{corlisssugar1982} and line data from \citet{fuhretal1988}. It was enhanced with oscillator strength data from \citet{thevenin1989,thevenin1990}.

We selected the well-studied and often observed $524.7$ nm, $525.0$ nm, $630.1$ nm,\ and $630.2$ nm\ lines for this work, which are often analyzed either individually or as line pairs \citep{stenflo1973, elmoreetal1992, tsunetaetal2008, solankietal2010, martinezpilletetal2011}. Altogether, the atom contained 23 continuum and 33 line transitions with a grand total of some 1300 wavelength points (we tried to keep this number small to reduce the computing resources needed). In this model atom, the oscillator strengths were as displayed in Table \ref{tab:f-values}. The $f$-values of the $524.7$ nm and $525.0$ nm lines were enhanced by roughly a factor of from two to three against those found in the literature \citep{blackwelletal1979}, following the correction procedure used in \citet{brulsvdluehe2001}. An iron abundance of 7.44 was employed following \citet{asplundetal2000b}. We note that we were mainly interested in differences between the line profiles obtained in LTE, 1-D NLTE, and 3-D NLTE under different conditions and not in the absolute line profiles. In particular, we did not compare our line profiles with observations, so that details such as the exact elemental abundance or $f$-values are not of central consequence and do not influence our conclusions.

The well-known problem of UV overionization \citep{brulsetal1992}, which leads to wrong population numbers because of missing iron line opacity in the UV, was eliminated by enhancing the opacities in the relevant wavelength range (opacity fudging) as proposed by \citet{brulsetal1992}. The resulting differences were calculated and are discussed in Sect. \ref{sec:fluxsheet_results_fudge}.

\subsection{Model atmospheres and computational setup}\label{sec:fluxsheet_model_atmos}
\begin{figure}
 \resizebox{\hsize}{!}{\includegraphics{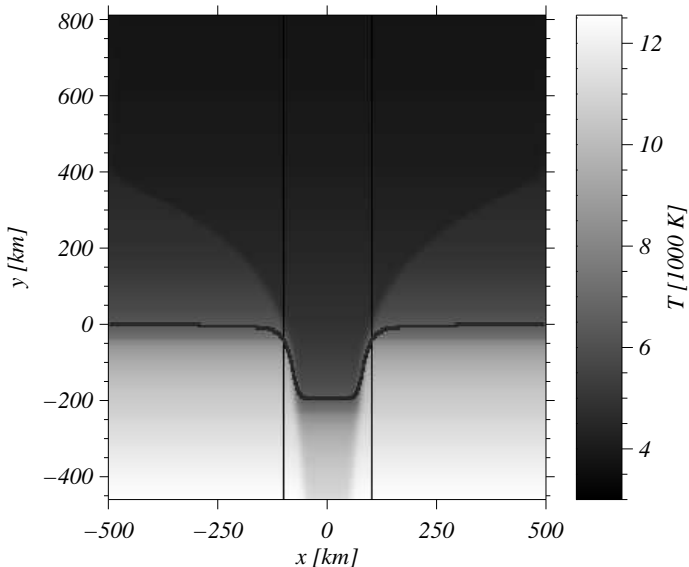}}
 \caption{ Temperature distribution in the atmospheric cross-section through a flux sheet model. The $\tau_c=1$ level (thick line) and the nominal boundaries of the flux sheet (thin vertical lines) at $\tau_c=1$ are indicated.} 
 \label{fig:atmos_temp}
\end{figure}
We studied both magnetic flux tubes (displaying axial symmetry) and flux sheets (displaying translational symmetry). In both cases, we employed the thin-tube approximation to describe their internal structure. For most computations, however, we used magnetic fields only to determine the Wilson depression and the expanding geometry of the flux tube, to better isolate the influence of multidimensional RT, as in \citet{stenholmstenflo1977}. In Sect. \ref{sec:fluxsheet_results_stokes}, however, we also consider results computed by including the formal solution of the Unno-Rachkowski equation, i.e. where we computed Stokes profiles using the so-called \emph{field-free} method \citep{rees1969}. Thus, we included the influence of the Zeeman effect on the line profiles (splitting, polarization), although the results presented here neglect the influence of the Zeeman effect on the population numbers.

The above-mentioned flux sheet could in principle be calculated with a 2-D code as long as one considers radiation emerging only along rays lying in a plane perpendicular to the plane of symmetry. However, to achieve an unbiased comparison with flux tubes, we used the 3-D code version. The qualitative effects in a 2-D flux sheet (FS) are the same as in a true 3-D flux tube (3-D FT), but are achieved at much lower computing costs. We note that an equidistant grid (in the $xy$-direction enforced by the code) is not very well-suited to mapping an expanding tube or sheet, therefore the number of grid points has to be quite high. To achieve reliable results, we used models with $288 \times 4 \times 256$ grid points in the flux sheet geometry, where the $z$-axis denotes the height in the atmosphere (see also Fig. \ref{fig:atmos_temp} for a cut across the flux sheet). In Sect. \ref{sec:fluxsheet_results_3d}, we present the results for a true 3-D flux tube geometry having $144 \times 144 \times 128$ grid points in the cartesian geometry of the RH code. Although only half the number of pixels are introduced along the $x$ and $z$ axes when compared to the flux sheet case, the total number of grid points is an order of magnitude larger. To keep the computing time from growing accordingly, the number of rays was reduced (standard set A6 of RH code instead of 54 vertical directions per octant, see Sect. \ref{sec:fluxsheet_angular_resol}). 

To compare our 3-D with our 1-D calculations, we constructed plane-parallel atmospheres by vertically cutting the 3-D cube at selected $x$-positions. The 1-D NLTE RH code was then run with the so-constructed 1-D atmospheres as input (we stick to the term 1-D NLTE for this method, which is often called 1.5-D NLTE). In addition, for comparison purposes 1.5-D computations in LTE were carried out for each flux sheet or tube (henceforth simply called LTE computations).

\subsubsection{Standard model}\label{sec:fluxsheet_model_atmos_std}
We varied a range of parameters such as radius, Wilson depression, thickness of the walls, filling factor, and flux sheet atmosphere (see Sect. \ref{sec:fluxsheet_results} for details). As a reference, we defined a standard model with a fixed parameter set:

{\it Atmosphere}: If not otherwise stated, the inner and outer atmospheres were analogous to those in \citet{stenholmstenflo1977} and therefore identical. This choice allowed us to more clearly segregate the effects of horizontal RT. We used the Harvard-Smithonian reference atmosphere (HSRA) described by \citet{gingerichetal1971}. In the field-free part, the model was extended downwards by combining it with the convection zone model of \citet{spruit1974}. Figure \ref{fig:atmos_temp} shows the geometry and the temperature distribution in the $xz$ plane for our standard model. The thick solid line follows $\tau_c=1$ and outlines the smooth but quite steep transition between the flux sheet interior and exterior. Inside the nominal boundary (indicated by the vertical lines as in many subsequent figures), the hot walls of the outer atmosphere, positioned below the continuum level of the quiet Sun, lie close to the local $\tau_c=1$ level, so that their presence strongly influences the radiation field sensed by an atom within the magnetic feature. A constant microturbulent velocity of $0.8$ km/s was added to every grid point of the model in direct correspondence with the original model. In the following, for simplicity, we refer to the outer, field-free atmosphere as either the quiet Sun (QS) or QS part. We note that the outer atmosphere ceases to exist at the height where the canopy fills the whole width of the atmosphere except for one single pixel at the edge of the atmosphere that was chosen to be pure HSRA without any magnetic field.

{\it Field strength}: The magnetic field strength of our standard sheet was $B=1600$ G at $\tau^{QS}_c=1$ and $B=0$ G outside, resulting in a Wilson depression of $191$ km. This B value cannot be directly compared with measured intrinsic field strengths, which depend on the height of formation of the employed spectral line(s), diagnostic method, etc. The field strength dropped with height and consequently the magnetic element expanded with height. We note that a flux sheet exhibits a more rapid expansion with height than a round flux tube as a consequence of flux conservation if one assumes a fixed run of field strength with height because $B \times R = const.$ for a sheet in contrast to $B \times R^2 = const.$ for a tube, where $R$ is the radius of the FT or the horizontal location of the FS boundary. 

{\it Wall thickness}: The sheet boundaries were smoothed in the $x$-direction with a Fermi function which drops from 90 to 10 percent within $20$ km, that is slightly above the minimal value of the wall thickness estimated by \citet{schuessler1986}.

{\it Radius, filling factor, and grid points}: Our standard model had a radius of $R=100$ km and a filling factor $\alpha=0.2$. The grid points were equidistant in each dimension, where $dx=3.47$ km, $dy=100$ km (irrelevant for a 2-D geometry), and $dz=5$ km, starting at $460$ km below $\tau_c=1$ of the field-free part of the model. Such a low starting point was needed to ensure that the model is optically thick at the bottom, even in the presence of a strongly evacuated flux sheet. 

\subsubsection{Angular discretization}\label{sec:fluxsheet_angular_resol}
\begin{figure}
\center{\resizebox{0.6\width}{!}{\includegraphics{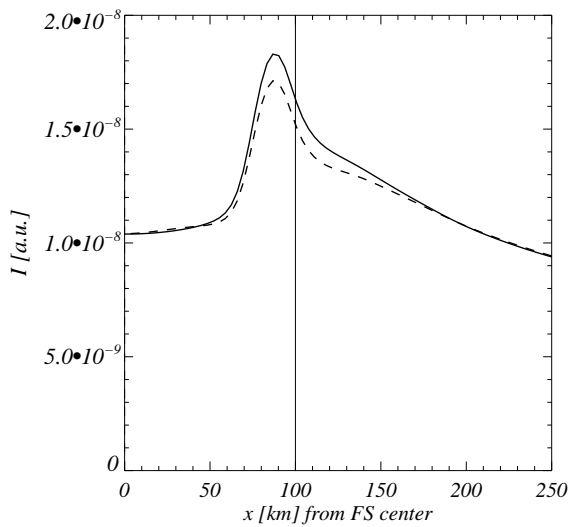}} }
 \caption{ 
 Vertically emergent intensity in the core of the $525.0$ nm line as a function of $x$-position, counted from the FS center (i.e. from the horizontal axis of Fig. \ref{fig:atmos_temp} cutting across the FS). The plotted curves were calculated with two different angle sets, where solid lines indicate the set of 27 (inclination) $\times$ 2 (azimuth) rays per octant and dashed lines the standard set A6 of the RH code with 6 rays per octant. The thin vertical line indicates the nominal FS boundary ($R=100$ km), which was reached at the $\tau_c=1$ level in the field-free atmosphere (at the same position as the rightmost vertical line in Fig. \ref{fig:atmos_temp}).
} 
 \label{fig:no_rays}
\end{figure}

Multi-dimensional problems basically need higher angular resolution to achieve the same accuracy of angular quadrature as  in a 1-D problem \citep[][]{vanNoortetal2002}. The very specific geometry of FS/FTs compelled us to increase the number of rays used in the simulation. We used a grid of 27 vertical (i.e. inclination) and 2 horizontal (i.e. azimuthal) angles per octant in the FS models, i.e. 216 outgoing rays from each grid point.  This small number of horizontal angles was adequate in the (2-D) FS geometry. Figure \ref{fig:no_rays} illustrates the difference in outgoing intensity between our chosen angular resolution and one obtained with the often used A6 standard set of the RH code with only 24 outgoing rays \citep[cf.][]{brulsvdluehe2001}. Close to the flux sheet boundary, the difference is significant ($\approx 7\%$). This difference has its origin in the very hot walls of the FS in this model and the optical thinness of the sheet's interior. The part where the walls are really hot is very localized, being restricted to a few grid points. Because the strongly evacuated sheet interior is optically thin, the hot walls can be seen from far away, as long as the ray travels only upwards through the sheet interior. The discretization of the rays gives rise to different lighthouse-like rays through the sheet. To see the hot walls at every point within the sheet it was necessary to increase the number of vertical rays. Some locations in the interior would not otherwise have been hit by a ray originating in the hot wall, which, in turn, may affect the local atomic level populations and the corresponding line depths. We note, however, that the artificially suppressed NLTE effects at these points are only relevant when the optical thickness of the interior is not negligible. 

For the 1-D NLTE calculation, we used a standard set of five Gaussian angles for all calculations.

\section{Results}\label{sec:fluxsheet_results}

\subsection{General features}\label{sec:fluxsheet_results_features}
\begin{figure}
\center{ \resizebox{1.0\width}{!}{\includegraphics{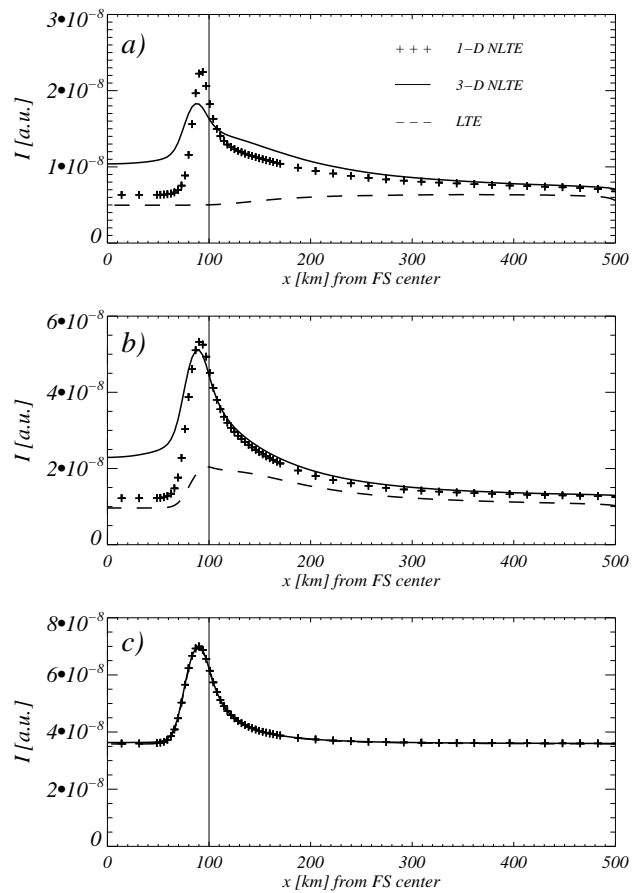}} }
 \caption{ 
 Intensities calculated with the different methods, indicated in the top panel, as a function of $x$ position at three different wavelengths in the $525.0$ nm line: a) line core; b) half depth ($\sim\!0.004$ nm); c) near continuum ($\sim\!0.008$ nm). The nominal FS boundary ($R=100$ km at $\tau_c=1$) is indicated by the vertical line. 
} 
 \label{fig:standard_Ix}
\end{figure}



\begin{figure*}
 \resizebox{\hsize}{!}{\includegraphics{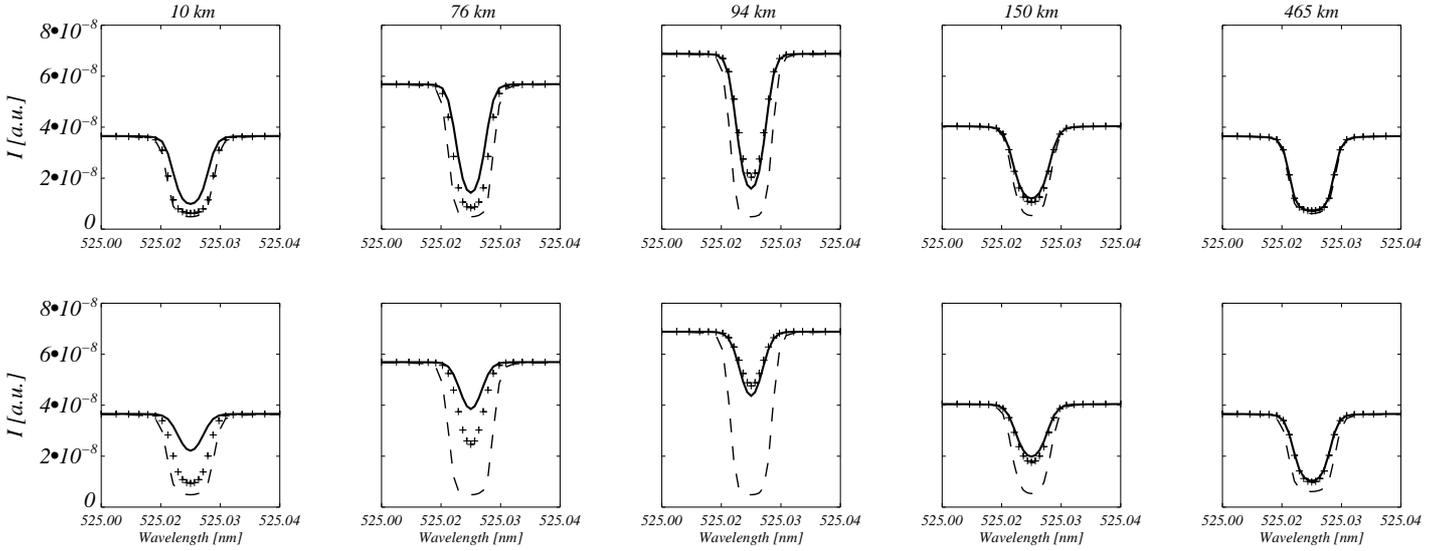}}
 \caption{ 
Intensity profiles of the $525.0$ nm line at five selected $x$ positions as indicated at the top of each panel (see also Fig. \ref{fig:standard_Ix}). Solid thick line: 3-D NLTE; dashed line: LTE; plus signs: 1-D NLTE. Upper panels: with opacity enhancement according to  \citet{brulsetal1992}; lower panels: no additional UV opacity.  
} 
 \label{fig:standard_opfudge_Inu}
 \label{fig:standard_opfudge_Inu1}
\end{figure*}

\begin{figure}
 \resizebox{\hsize}{!}{\includegraphics{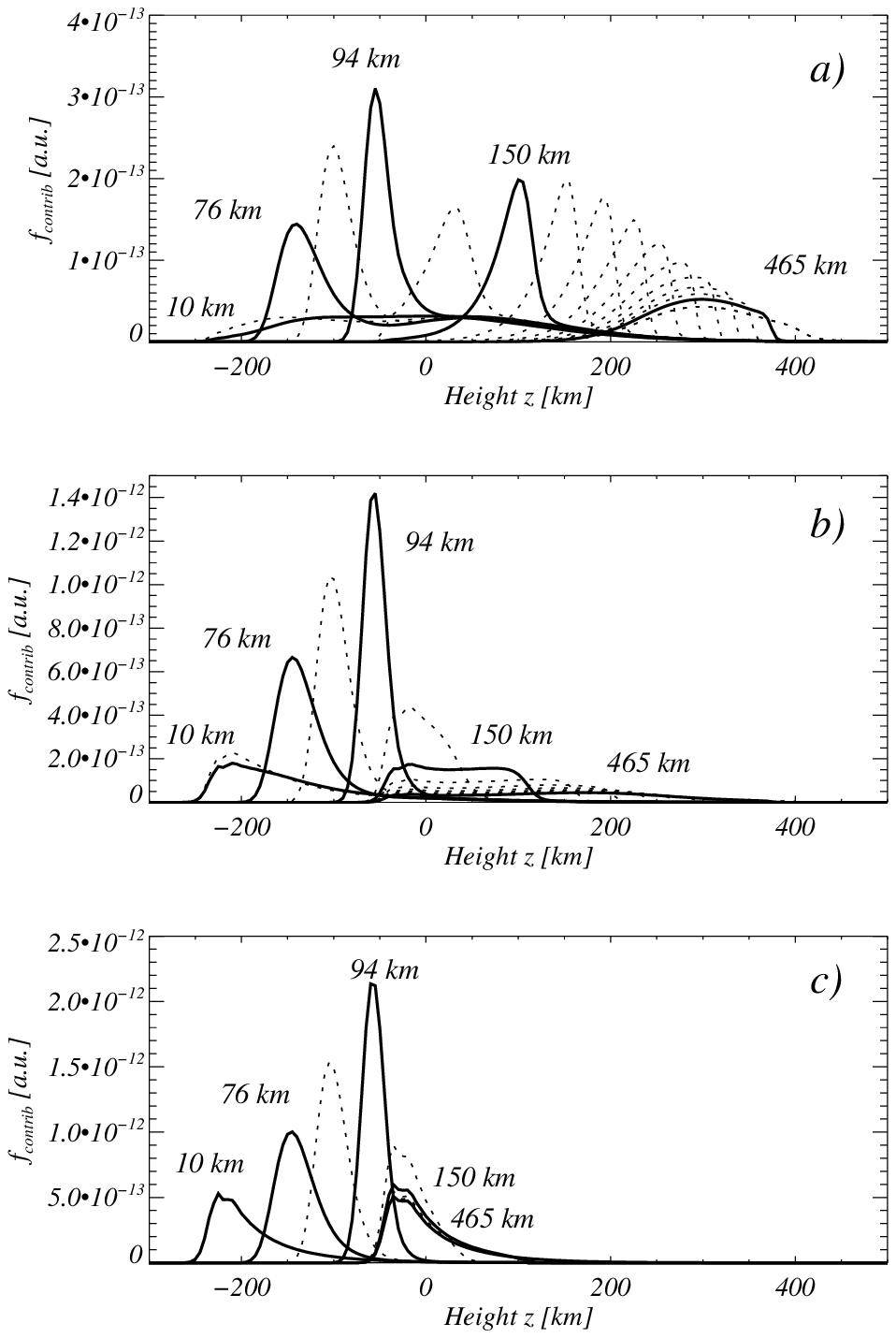}}
 \caption{ 
Contribution functions computed in 3-D NLTE at different $x$-positions for the same three selected wavelengths of the $525.0$ nm line as in Fig. \ref{fig:standard_Ix}: a) line core, b) line flank, c) continuum. The dotted contribution function is given for every tenth grid position corresponding to $\Delta x \approx 35$ km, starting at $10$ km from the plane of symmetry of the flux sheet. The five contribution functions  corresponding to the same $x$-positions as the profiles shown in Fig. \ref{fig:standard_opfudge_Inu1} are highlighted (solid lines) and indicated by their $x$-position. Note the very different shapes and positions of the individual curves. 
} 
 \label{fig:standard_contrib}
\end{figure}

We like to compare our results for the LTE, 1-D NLTE, and 3-D NLTE RT computations. We first concentrate on the \ion{Fe}{I} $525.0$ nm line formed in our standard atmosphere (see Sect. \ref{sec:fluxsheet_model_atmos_std}), but later also consider the other spectral lines and models.  Figure \ref{fig:standard_Ix} shows the intensity along the $x$-axis of the standard flux sheet (i.e. cutting across the sheet) for three selected wavelength positions, namely in the line core, in the line flank at approximately half depth ($\sim\!0.004$ nm from the line center), and near the continuum ($\sim\!0.008$ nm from the line center). The nominal flux sheet boundary at $\tau_c=1$ is indicated by vertical lines. The center of the flux sheet is located at $x=0$. Only half of the cut through the sheet is plotted since the other half is simply a mirror image owing to the symmetry of the model. 

In the line core and flank, both 1-D and 3-D NLTE computations show similar, strong variations along $x$ (Fig. \ref{fig:standard_Ix}), as reported by \citet{brulsvdluehe2001}, with the rest intensity reaching a peak value of two to four times that found in the QS far away from the FS. We note that quantitative differences between the results of \citet{brulsvdluehe2001} and ourselves may be caused by the quite different atomic and atmospheric parametrizations, the most important probably being the Wilson depression (magnetic field strength). In the line wing, the variation is slightly higher than in the line core. The LTE intensity, however, shows only a minor variation along the $x$-axis in both the line wing and core. Far away from the FS, all three methods of computation give almost identical results, only the LTE rest intensity remaining slightly below that of the NLTE methods. In the continuum, all three methods are identical and show a sharp rise in the rest intensity close to the boundary of the FS \citep[cf.][]{caccinseverino1979}. 

Finally, a large difference between the 1-D and 3-D NLTE calculations is found inside the FS, with the 3-D profile being weakened (i.e. having a higher rest intensity than the profiles computed with the other methods).

The upper panels of Fig. \ref{fig:standard_opfudge_Inu1} show the corresponding spectra for five selected locations indicated above the panels. In all plotted cases, the LTE line profile saturates, maintaining nearly the same rest intensity (as seen in Fig. \ref{fig:standard_Ix}a, while the NLTE profiles become much weaker at some locations. We note that the relative differences in the line depth (see Fig. \ref{fig:standard_opfudge_Inu1}) are considerably smaller than those in rest intensity (Fig. \ref{fig:standard_Ix}) owing to the generally low line core intensity.

We now attempt to understand the intensity variations plotted as a function of $x$ in Fig. \ref{fig:standard_Ix}. Far away from the flux sheet, the line is formed in the QS part of the atmosphere just below the optically thin canopy, which lies at a height of $400$ km. The difference between the LTE and NLTE profiles at that location is basically the same as one gets when using the HSRA on its own, without any FS. As we move closer to the sheet, the height of the canopy's base decreases continually. Unit optical depth in the line is reached at lower heights (below the canopy and, for a sufficiently low canopy base, at its boundary) owing to the lower density and hence opacity of the canopy, which remains optically thin to the line right up to positions slightly inside the nominal FS boundary. The line core is therefore formed in the hotter layers of the external atmosphere and consequently has a higher line core intensity, which culminates just inside the sheet's nominal boundary. 

The 3-D NLTE contribution functions in Fig. \ref{fig:standard_contrib}a  illustrate the decreasing height of the line-core formation region as we move towards the sheet. We note the peculiar shape of the contribution functions far away from the sheet. In contrast to the typical case in which the contribution function first sharply increases when moving up in the atmosphere, then slowly decreases above the height of maximum contribution, these contribution functions show a steep edge at the higher end, which corresponds to the lower boundary of the canopy. Examples are found between the two highlighted contribution functions (solid lines) at $x=465$ and $x=150$ km.

The position of this edge moves down as we move closer to the sheet in complete accordance with the geometry of our atmosphere (see also Fig. \ref{fig:atmos_temp}). The intensity maximum of $I_{core}$ for the NLTE-cases lies slightly inside the flux sheet, where we---looking down vertically---directly see the hot walls below the $\tau_c=1$ level of the QS part. All three contribution functions (at the line core, at half-depth, and in the continuum, i.e. Fig. \ref{fig:standard_contrib}a, \ref{fig:standard_contrib}b, and \ref{fig:standard_contrib}c, respectively). At that position, the contribution functions are peaked almost exactly at the same height, confirming the hypothesis that we see light formed below the height at which the continuum is formed far away from the FS. We note that the spectral line does not completely disappear owing to a small difference between the line and continuum formation levels and the very strong temperature gradient at the hot walls.

Why do we not see the same effect in the line formed under LTE? As the population numbers of \ion{Fe}{i} are generally higher in LTE (owing to missing ionization by UV radiation), the line is slightly stronger than in the NLTE case. This leads to the situation where the canopy starts to become optically \emph{thick} in the line core even far from the FS. Closer to the FS, the optical thickness of the canopy increases as the height of its base decreases. Because the temperature in either the canopy or the FS at heights of around $300-400$ km is only slightly cooler than in the external atmosphere at the same geometrical height, the line core intensity decreases only slowly when we move towards the sheet (i.e. the line becomes slightly stronger). 

Moving away from the line core, the opacity decreases, so that in the LTE continuum the intensity shoots up, as in the NLTE line cores (Fig. \ref{fig:standard_Ix}c ). The three identical continuum curves in Fig. \ref{fig:standard_Ix}c indicate that the continuum is as expected formed in LTE. In the line flanks, the LTE opacity takes on an intermediate value so that the intensity still increases when horizontally approaching the FS boundary until reaching a gentle maximum at the FS boundary (see Fig. \ref{fig:standard_Ix}b ).

The main effect of 3-D RT is seen within the sheet. There the LTE and 1-D NLTE profiles attain nearly the same depth as they have far from the flux sheet, as expected for a model with the same atmospheres inside and outside the flux sheet. Inside the flux sheet the LTE and 1-D NLTE profiles are independent of $x$ because the local atmosphere does not change (when neglecting the wall smoothing). The 3-D NLTE line profile, however, remains significantly weaker (cf. Fig. \ref{fig:standard_opfudge_Inu1}e) owing to the ionization of neutral iron by the UV radiation flowing in sideways from the hot walls of the flux sheet, as discussed by \citet{stenholmstenflo1977}. The line-core contribution function becomes very broad, extending from the continuum forming level at $-200$ km up to $+300$ km in the sheet. 

In a small region at the FS boundary, the 1-D NLTE intensity exceeds the 3-D case. In the hot walls, both the horizontal RT towards the cool sheet interior and radiation from the cool interior of the flux sheet (which are both not present in the 1-D case) decrease the amount of UV radiation reaching an atom there, thus giving rise to the opposite effect with respect to that seen inside the flux sheet.

We conclude this section by restating \citet{stenholmstenflo1977} since they provided a succinct summary of the results presented so far: ``Even when a line in a plane-parallel atmosphere is formed \emph{in LTE} (i.e. LTE is a reasonable approximation), there is no guarantee that LTE is still a good choice in more complex geometrical configurations.''

\subsection{The role of UV opacity}\label{sec:fluxsheet_results_fudge}
It is important to judge the influence of enhancing the UV opacity following \citet{brulsetal1992} to compare with the work of \citet{stenholmstenflo1977}, who used a simple two-level atom without any enhanced UV opacity. The role of UV opacity is to partially suppress the influence of the UV radiation from the deeper, hotter layers and therefore reduce the NLTE effect of UV overionization.

The spectra in the lower panels of Fig. \ref{fig:standard_opfudge_Inu} are calculated without any enhanced UV opacity. Even far away from the sheet, the NLTE-profiles differ rather significantly from the LTE profiles when no opacity enhancement is introduced. They start to differ dramatically when approaching the sheet (at $150$ km). The difference between the 1-D and the 3-D calculations in the sheet is also strongly  enhanced due to the strong influence of the horizontal transport of UV radiation in the 3-D case. The depth of the line in the 3-D case is smaller by more than a factor of two. These results are more comparable to those obtained by \citet{stenholmstenflo1977}. Obviously the employment of a more adequate UV opacity model reduces not just the strength of NLTE effects in general, but in particular the effect of the horizontal RT, which remains important nontheless inside the magnetic element.

We note that the influence of the UV opacity enhancement on the \ion{Fe}{i} $630.1$ nm line, which is principally more sensitive to NLTE effects, in particular horizontal transport, is slightly smaller than that on the $525.0$ nm line. The $630.1$ nm line is already in a partially saturated regime and therefore removing the UV-opacity enhances the NLTE effects to a lesser extent.

\subsection{Behaviors of different lines}\label{sec:fluxsheet_results_lines}
\begin{figure*}
 \resizebox{\hsize}{!}{\includegraphics{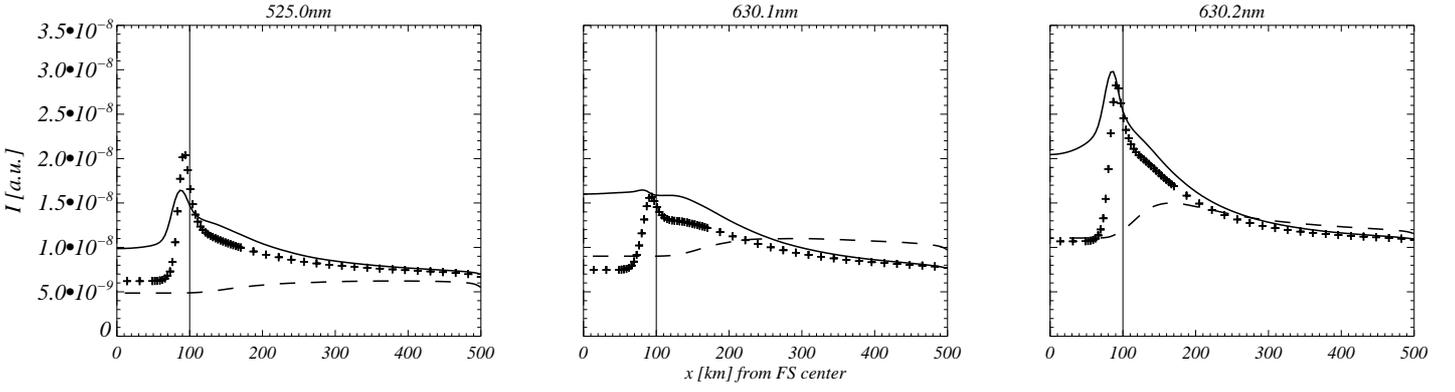}}
 \caption{ 
 Line-core intensity of three spectral lines (identified above each frame) along the $x$-axis of the model. Solid: 3-D NLTE, dashed: LTE, plus signs: 1D-NLTE. 
} 
 \label{fig:standard_3l}
\end{figure*}

\begin{figure}
 \resizebox{\hsize}{!}{\includegraphics{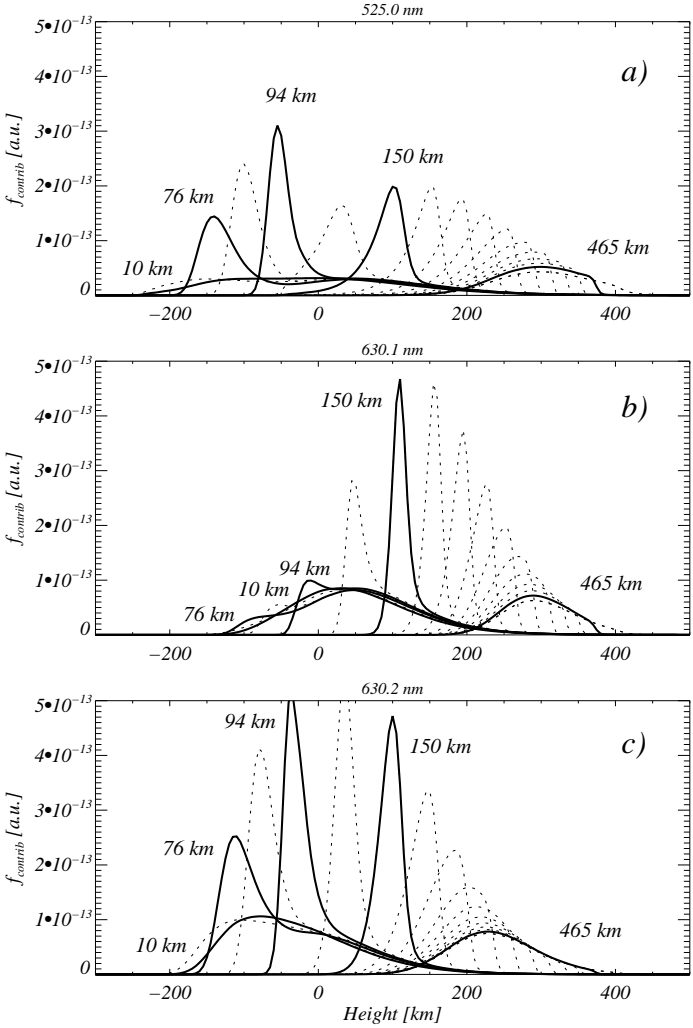}}
 \caption{ 
Line-core contribution functions as in Fig. \ref{fig:standard_contrib}a for the three lines $525.0$ nm (a), $630.1$ nm (b),  and $630.2$ nm (c).
} 
 \label{fig:standard_3l_contrib}
\end{figure}

The discussion in Sect. \ref{sec:fluxsheet_results_fudge} already indicates that the effects of NLTE and 3-D RT depend strongly on the formation height, so that different lines may exhibit different behaviors. In this section, we compare the profiles of three widely employed lines (\ion{Fe}{i} $525.0$ nm, \ion{Fe}{i} $630.1$ nm, and \ion{Fe}{i} $630.2$ nm). The results for the $524.7$ nm line  turn out to be exactly the same as for the $525.0$ nm line, which is unsurprising, given that the two lines are nearly identical, save their Land\'e factors (which do not play a role here, since we set $B=0$ G for the line-formation computations). The $630.1$ nm line is expected to be somewhat stronger, and the $630.2$ nm line to be somewhat weaker than the $525.0$ nm line. In addition, the two $630$ nm lines have a considerably higher excitation potential, which may also influence our results.

Figure \ref{fig:standard_3l} shows the line core intensities along the $x$-direction as in Fig. \ref{fig:standard_Ix}a, but now for the three lines $525.0$ nm, $630.1$ nm, and $630.2$ nm. We repeat the results for the $525.0$ nm line from Fig. \ref{fig:standard_Ix}a for ease of comparison. The global features are the same for all three lines. Far away from the sheet, the lines are formed in a very similar way (although the $630$ nm doublet is somewhat stronger in NLTE relative to its strength in LTE). Close to the flux sheet boundary, there are, however, notable differences.

A major difference of the $630.2$ nm line in comparison to the other lines is the generally higher intensity level owing to its weaker opacity. The LTE curve confirms this view, as can be deduced from the increasing $I_{core}$ level as we move towards the sheet: this is because the LTE line is formed (partially) in the hotter gas below the canopy and not mainly either in the canopy itself or its boundary where the cores of the other lines are formed (see also the discussion in Sect. \ref{sec:fluxsheet_results_fudge}). The significantly stronger 3-D effect seen in this line is quite striking inside the FS compared to that for the $525.0$ nm line. 

The 3D-NLTE case of the $630.1$ nm line differs quite strongly from those of the other lines. Although the intensity increases towards the sheet, there is no peak near the flux sheet boundary, which remains at almost the same level over the whole sheet. The line-core contribution functions drawn for the three lines in Fig. \ref{fig:standard_3l_contrib} indicate that the weaker lines have strong contributions from the continuum near the boundary (at $x=94$ km), in contrast to the stronger $630.1$ nm line. The contribution function in the line core is clearly shifted towards higher levels, i.e. we do not see directly the hottest parts of the walls. We note that  outside the sheet (e.g. at $x=150$ km), the $630.1$ nm line is formed only marginally above the other lines. Inside the sheet, the effect of horizontal RT weakens this line, leading to higher intensities in the 3D-NLTE case. 

Figure \ref{fig:standard_3l}b indicates that at certain $x$-positions the LTE intensity of the $630.1$ nm line is slightly higher than in NLTE. This may be explained by mainly two different effects (Jo Bruls, private communication).  The photospheric source functions for these lines usually follow the Planck curve for both LTE and NLTE. In the NLTE case, owing to the lower population of Fe I, the $\tau = 1$ level is shifted downwards, into hotter regions, weakening the line. On the other hand--especially for the stronger $630.1$ nm line formed higher up--the NLTE source function may clearly fall below the Planck function, leading to the opposite effect. Which effect wins in the end is determined by the complex interplay of many transitions and levels in the atom and is difficult to forecast.

\subsection{Influence of the wall thickness}\label{sec:fluxsheet_results_smooth}
\begin{figure}
\center{\includegraphics[width=0.35\textwidth]{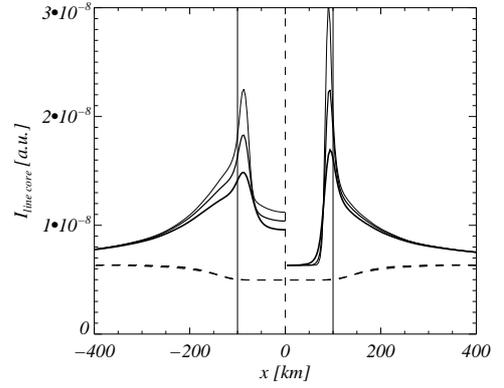}}
 \caption{ 
 Line-core intensity of the \ion{Fe}{i} $525.0$ nm line along the $x$-axis for three different atmosphere smoothings (thick curves $40$ km; normal $20$ km; thin $10$ km). 3-D NLTE solid curves on the left; 1-D NLTE solid right; LTE dashed.
} 
 \label{fig:standard_smoothing}
\end{figure}

The true thickness of the walls of magnetic elements is not well-known. The minimum thickness based on the expected solar magnetic resistivity, diffusivity etc. was estimated by \citet{schuessler1986} to be on the order of $3-10$ km. In addition, radiation is expected to smooth the transition from the surroundings to the FT/FS and convection is expected to make the wall irregular, although the profile shape of the $g=3$ line at $1.56 \mu$m suggests that the magnetic field drops rather suddenly at the edge of the magnetic element \citep{zayeretal1989}. In our calculations, we used smoothings with three different Fermi functions, each defined by the distance over which one atmospheric component changes from the 90\% to the 10\% level. In total, thicknesses of $40$ km, $20$ km (standard model), and $10$ km were tested. We expected the wall thickness to have a strong influence on the lines where the continuum radiation of the hot wall could be seen directly, as in the case of the $525.0$ nm or the $630.2$ nm lines. By changing the distance over which the atmosphere changes from the cold interior to the hot QS outside the sheet, we directly changed the temperature at the locations where the continuum radiation forms. We also expected that the effect of horizontal RT would decrease when we expanded the thickness of the flux sheet boundary, since a smoother transition brings the FS ever closer to the plane-parallel, i.e. the 1-D case. The resulting intensity is determined by a complex interplay of many factors \citep[see also ][]{brulsvdluehe2001}.

Figure \ref{fig:standard_smoothing} shows the LTE, 3-D NLTE, and the 1-D NLTE line-core intensities of the $525.0$ nm line along the $x$-axis for the three different wall thicknesses. As expected, the strongest influence is seen at the hot wall itself. While the LTE curves almost coincide, the 1-D and 3-D NLTE curves show a systematic increase in the intensity maximum with decreasing wall thickness. Interestingly, the change of the 3-D intensities in the inner part of the sheet is, however, much smaller. The largest effect is displayed by the $524.7$/$525.0$ nm lines. The other lines (not shown), especially the $630.1$ nm line, are less sensitive to the wall thickness.

\subsection{Variation in the radius }\label{sec:fluxsheet_results_R}
\begin{figure}
 \center{\includegraphics[width=0.35\textwidth]{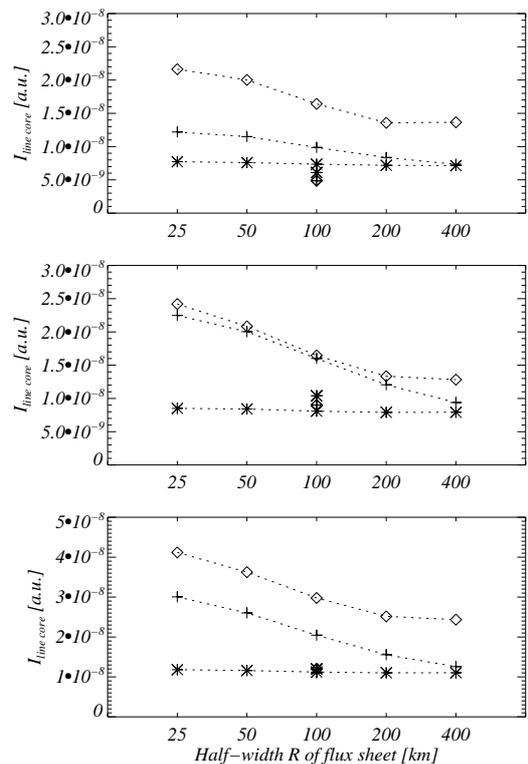}}
 \caption{ 
 Intensity in the line core of the three lines ($525.0$ nm upper, $630.1$ nm middle, and $630.2$ nm lower panel) at selected $x$ positions (diamonds maximum intensity near boundary of FS, plus signs FS center, asterisks QS far away from FS) for models with different half-widths R of the FS. At $R=100$ km, we have added the corresponding LTE values, which do not vary with FS width. 
} 
 \label{fig:var_r_constgeom}
\end{figure}

We next considered FS models with different widths. To change the width, the flux sheets were stretched in the $x$ direction. Therefore, the thickness of the walls scaled with the width of the FS. All other quantities remained the same. 

Core intensities at three locations along the $x$-axis are plotted in Fig. \ref{fig:var_r_constgeom} versus the FS half-width of three spectral lines. The selected locations are the sheet center (plus signs), the position of the intensity maximum near the wall of the FS (diamonds), and the QS far away from the sheet (asterisks). The latter does not show much variation with radius as a consequence of its position far away from the sheet. For $r \ge 400$ km, the intensity at the FS center almost corresponds to the 1-D NLTE case. As expected, these broad flux sheets are optically thick in the horizontal directions, so that the hot walls are no longer visible at the center of the FS, hence do not influence the radiation field there. 

The intensity maximum near the boundary (diamonds) displays a similar behavior. This can be understood by bearing in mind that the wall thickness increases with FS width, so that for wider tubes the walls are not so hot (see Sect. \ref{sec:fluxsheet_results_smooth}). However, horizontal transport plays a role as well, since an observer located close to one wall also sees radiation from the opposite wall. With increasing FS half-width, the strength of this irradiation decreases. For the intensity maximum at the flux sheet boundary, the effect of horizontal RT seems to saturate at $R \approx 200$ km. This is consistent with the behavior seen at the center of the flux sheet where the effect saturates at double the distance, i.e. at $R \approx 400$ km. Therefore, the opposite wall (at 2R) is invisible at the boundary and can no  longer influence the line for broader FS. 

A few test computations with FS of different half-width, but with the same wall thickness also show a decrease in the core intensity with FT size. The decrease is only slightly less steep than in Fig. \ref{fig:var_r_constgeom}, confirming that the main effect is due to the FS size and not due to the wall thickness, as already suggested by a comparison of Fig. \ref{fig:var_r_constgeom} with Fig. \ref{fig:standard_smoothing}.

\subsection{Variation in the temperature stratification }\label{sec:fluxsheet_results_atmos}
\begin{figure}
\center{\includegraphics[width=0.45\textwidth]{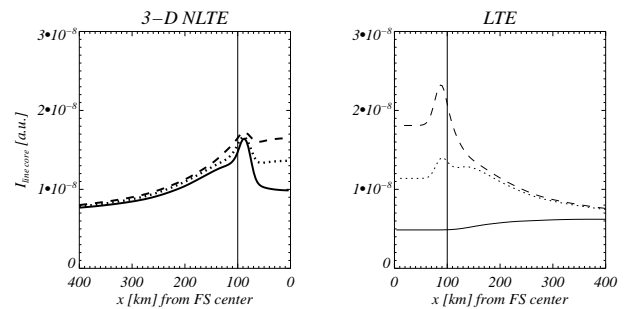} }
 \caption{ 
The $525.0$ nm line-core intensities (3-D NLTE left, LTE right) along the $x$-axis for three different FS atmospheres (solid HSRA (standard), dashed NET, dotted PLA). In the field free part, we always use HSRA.
} 
 \label{fig:atmos_var}
\end{figure}

We then modified the model by replacing the atmosphere in the FS \emph{interior} by, in turns, the network model of \citet{solanki1986} (from now on referred to as NET) and a plage model according to \citet{solankibrigljevic1992} (hereafter PLA). The nominal field strength of the FS ($B=1600$ G) and the external QS atmosphere (HSRA) were unchanged. Although these semi-empirical models are relatively old, they agree well with more recent 3-D radiation MHD simulations \citep{vitasetal2009}. The NET model is the hottest among the three selected atmospheres, being hotter than both the HSRA and the PLA model at all $\tau$. The PLA model is similar to the HSRA at heights around $\tau_c=1$, but has a flatter temperature gradient and is therefore hotter than the HSRA at the formation height of the $525.0$ nm line core, which was taken into consideration here (see Fig. \ref{fig:atmos_var}).

Far away from the FS, the rest intensities are similar for all models and computational methods, which is unsurprising as the canopy lies so high that the FS atmosphere is almost irrelevant. Close to or inside the FS, however, the differences are significant for both the different models and methods.

We first considered the LTE case, which is easy to understand because the different temperatures of the inner atmospheres are directly visible in the resulting rest intensities. The closer to the FS we are, the more we see the influence of the FS atmospheres. In accordance with the higher temperature in the PLA and NET models, the rest intensities increase towards the FS (right panel in Fig. \ref{fig:atmos_var}). The highest intensity is reached near the FS boundary and for the hottest model (NET). The weakening of \ion{Fe}{i} lines in hotter atmospheres is due to the enhanced ionization of iron.

In the NLTE case, the rest intensities vary strongly as well, but only in the interior of the FS. For the NET model, the NLTE intensity lies even below the LTE value. This can be explained by the large temperature difference between the NET model and the HSRA at the geometric height of line core formation. The downward shifted NET model in the FS is still approximately 200 K hotter than the external HSRA model. We note that this no longer applies at greater depth. In the sheet, the LTE line-core intensity is therefore even higher than the corresponding 3D NLTE intensity because the flux sheet interior is "cooled" by horizontal RT towards the now "cold" walls. The usual line weakening by the horizontal transport of UV radiation, as seen, e.g., in the HSRA-HSRA reference case, is inverted and the net horizontal transport of radiation has the opposite sign. 

The surprisingly good agreement between the PLA-LTE and the PLA-3D-NLTE rest intensity leads to the  conclusion that the PLA model--when calculated in LTE--compensates the most effectively for 3D and NLTE effects. The other lines considered in this paper also agree quite well (not plotted). This suggests that the PLA model of \citet{solankibrigljevic1992} also remains valid as a description of the atmosphere of magnetic elements in full 3-D NLTE, at least for the spectral lines considered here.

\subsection{Influence of geometry: Flux tubes}\label{sec:fluxsheet_results_3d}
\begin{figure}
\center{ \includegraphics[width=0.5\textwidth]{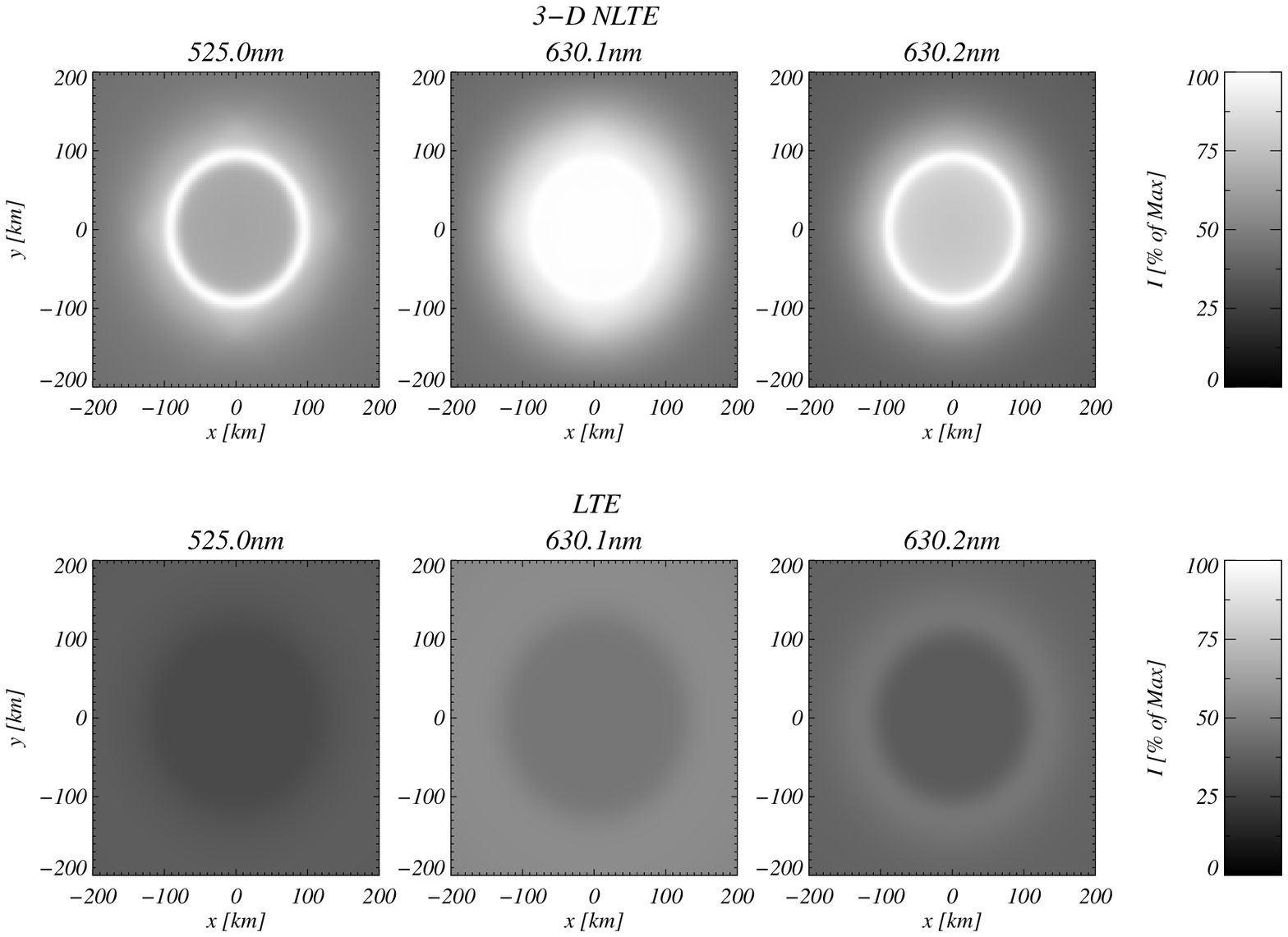}}
 \caption{ 
 Intensity maps of a flux tube model in the cores of the three lines indicated above each panel. The grayscale for the 3-D NLTE (upper panels) and LTE (lower panels) cases are the same within the pair of plots for a particular spectral line. The scale varies, however, from one spectral line to another between zero intensity and the maximum achieved in 3-D NLTE (upper panels), which is set to 100\%. 
} 
 \label{fig:ft3_tv}
\center{ \includegraphics[width=0.45\textwidth]{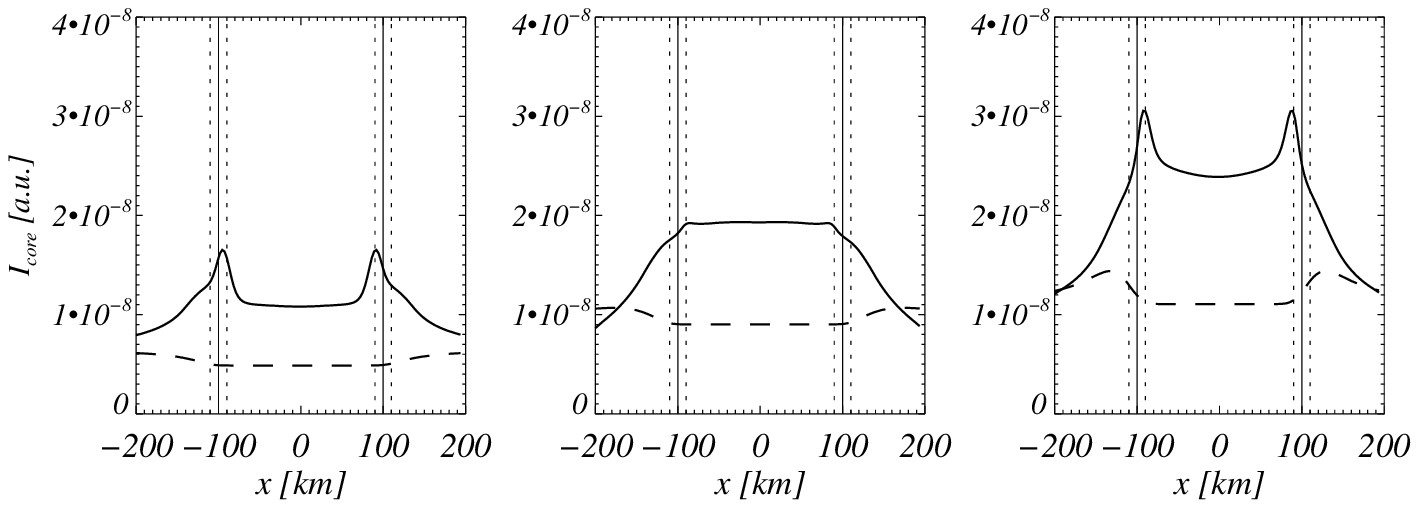}}
 \caption{ 
 Intensity in three line cores (from left to right: $525.0$ nm, $630.1$ nm, $630.2$ nm) along a slice passing through the center of the rotationally symmetric flux-tube (FT) model. Vertical lines indicate the FT boundaries (solid) and the wall thickness (dashed). Solid lines 3-D NLTE; dashed lines LTE. 
} 
 \label{fig:ft3_Ix}
\end{figure}


We present results for a flux tube (FT) geometry, i.e. for magnetic elements with a round cross-section. Despite the lower   resolution of the atmospheric model (cf. Sect. \ref{sec:fluxsheet_model_atmos_std}), the tube may be represented in a fairly smooth way because the filling factor of 0.2 leads to a considerably larger FT fraction when we look along only one dimension (e.g. $x$-axis). The hot walls are therefore resolved almost as well as in the FS case. All other model ingredients (atomic model, atmospheric components, radius, magnetic field, filling factor, and smoothing) remained unchanged from our standard FS model. 

Fig. \ref{fig:ft3_tv} shows maps of the core intensities of three lines computed in 3-D NLTE (top) and LTE (bottom). The grayscales for the 3-D NLTE and the corresponding LTE images are the same for a given spectral line to facilitate the comparison. Owing to the rather different maximum intensities reached by the different lines, the greyscales were normalized individually. The maximum intensity per line---typically reached near the FT boundary in the 3-D NLTE case---was used as the reference value (100\%). Fig. \ref{fig:ft3_Ix} displays the same intensities along a cut passing through the axis of the FT. We observed the same effects as for the standard flux sheet, although the line core intensities in the tube itself are higher than for the FS, which may be explained by the more intense irradiation from the now completely surrounding hot walls. Differences near the FT boundary originated from the different geometrical expansion of the FT in three dimensions. 

This confirms that the results obtained for the (2-D) FS geometry apply qualitatively to the 3-D FT case as well. The horizontal transfer effects, i.e. those of 3-D RT, are actually larger in FTs than for a FS (see e.g. the line weakening of the $525.0$ nm line at the centers of the FT and FS).

\subsection{Center-to-limb variation }\label{sec:fluxsheet_results_clv}
\begin{figure*}
\center{{\includegraphics[width=0.75\textwidth]{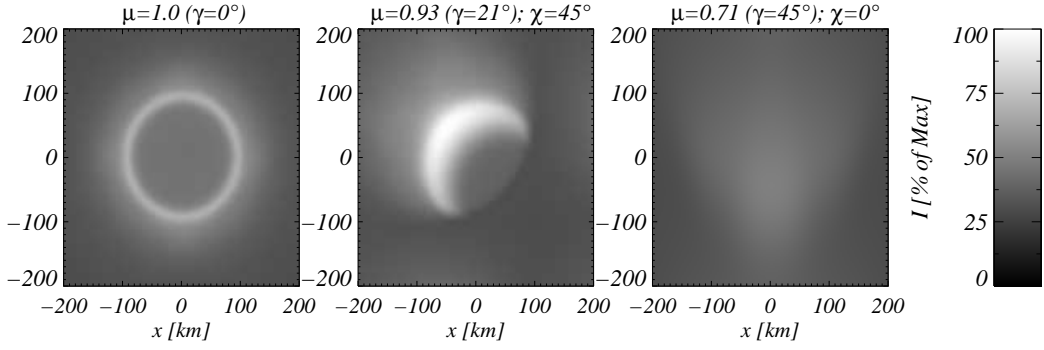}} }
 \caption{ 
 Intensity map of the 3-D FT in the core of the $525.0$ nm line for three different viewing directions as indicated on top of the panels. See Sect. \ref{sec:fluxsheet_results_clv} for definitions of $\gamma$ and $\chi$. The maximum intensity of the second image, showing the highest intensity values, was used to normalize the grayscale of all images. 
} 
 \label{fig:ft3_tv_3ang}
\end{figure*}

\begin{figure}
\center{{\includegraphics[width=0.5\textwidth]{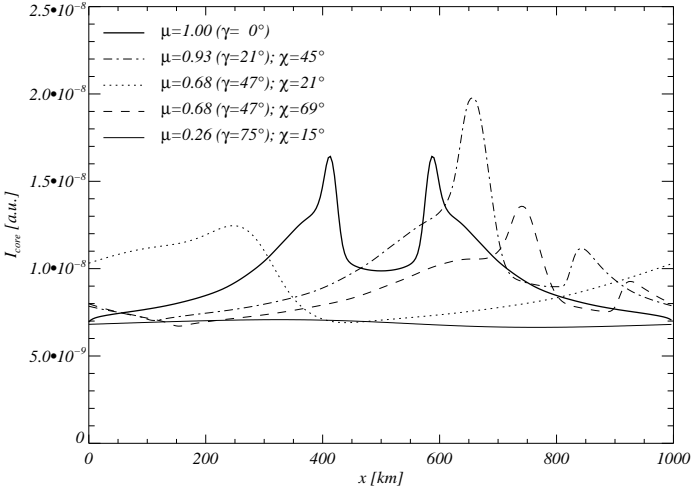}} }
 \caption{ 
 Core intensity of the $525.0$ nm line for the different viewing angles considered in  the 2-D FS geometry. Viewing angles and $\mu=cos \gamma$ are given in the text labels. In a FS, rays with the same $\mu$ (dotted, dashed) do not necessarily have similar intensities. The hot walls are more clearly seen for larger azimuthal angles $\chi$. Note that the location of the maximum intensity in $x$ is---for inclined rays---somewhat arbitrary as the outgoing intensity is plotted at the top of the atmosphere.  
} 
 \label{fig:ft_divang}
\end{figure}

Fig. \ref{fig:ft3_tv_3ang} depicts the synthetic images of a flux tube in the intensity (3D-NLTE calculation) of the $525.0$ nm line core for three different viewing angles. In the following, $\gamma = \arccos\rm\mu$ denotes the angle between the vertical ($z$-axis) and the line of sight (LOS) and $\chi$ the azimuthal angle in the $xy$ plane measured from the $x$-axis. In the leftmost panel, showing the vertical view (i.e. $\mu=1.0$), we see the strong and very localized ($\approx 10-20$ km) intensity maximum near the FT boundary and, less strikingly, but still clearly visible, the enhanced intensity in the interior of the FT. For $\mu=0.93$ (i.e. $\gamma=21.5^{\circ}$) and $\chi=45^{\circ}$ (middle panel), we still see half of the ring, which is now much broader ($\approx 50$ km) and brighter. It actually resembles an arched structure such as a small loop.  The rest of the hot wall is hidden from view by the optically thick cooler gas of the external atmosphere in the foreground. For $\mu=0.7$ ($\gamma=45^{\circ}$) and $\chi=0^{\circ}$ (right panel), the hottest parts of the wall cannot be seen directly anymore, leading to a more diffuse intensity distribution, partially produced by the warm walls at greater height, partially due to the horizontal RT. We note that for all LTE calculations, no significant structures could be recognized at all. The same is true for NLTE computations at $\mu<0.7$ ($\gamma>45^{\circ}$). 

Fig. \ref{fig:ft_divang} shows the intensity distributions for the same (and more) viewing angles, but now for the 2-D flux sheet. In the FS geometry, there may be different intensity distributions for a single $\mu$ and different azimuthal angles, as illustrated for $\mu=0.68$. When the ray is mainly in the $x$-direction, i.e. $\chi$ is small (dotted curve), we see the same smooth distribution as for the 3-D FT (rightmost panel of Fig. \ref{fig:ft3_tv_3ang}). If we look at the same $\mu=0.68$, but now mainly along the $y$-direction, i.e. $\chi$ is large (dashed curve), we have a more direct view of one of the hot walls because the ray is oriented along the central plane of the flux sheet. Figure \ref{fig:ft_divang}  shows an additional viewing angle with a very small $\mu$ of 0.26 ($\gamma=75^{\circ}$, thin solid line), where one barely recognizes any variation, since the hot wall is almost completely hidden by the front part of the FS.

\subsection{Solution for the full Stokes vector}\label{sec:fluxsheet_results_stokes}

\begin{figure*}
\center{{\includegraphics[width=0.65\textwidth]{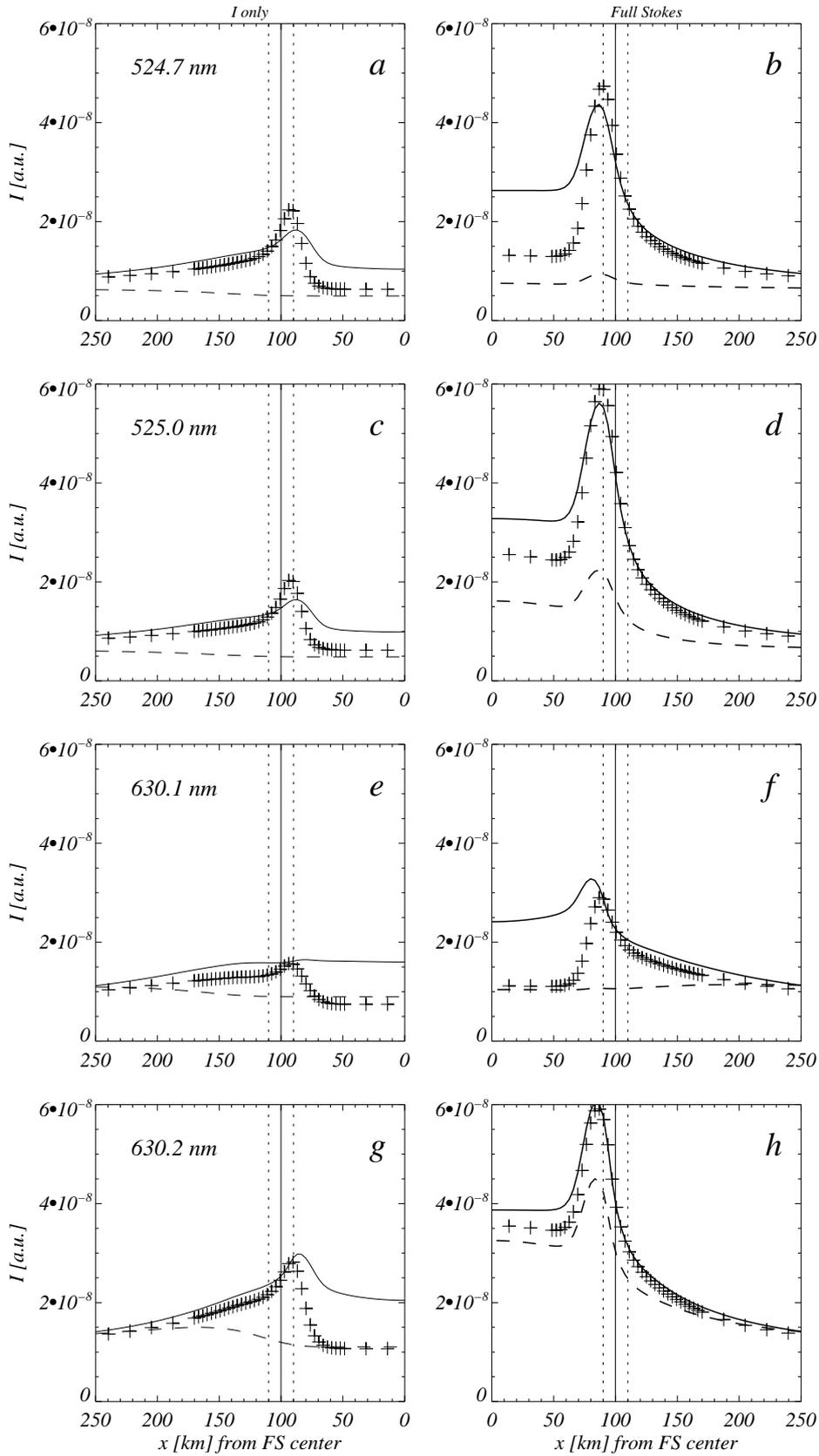}} }
 \caption{ 
Core intensities for the four selected lines and different calculation methods (solid 3-D NLTE, dashed LTE, and plus signs 1-D NLTE; the vertical lines are as in previous figures). Each panel pair shows the intensities of a single line for a formal solution with Stokes $I$ only (i.e. no Zeeman splitting; panels a, c, e, g) and one considering the full Stokes vector ( panels b, d, f, h) as indicated above each panel. Note the $x$-scale in the left panels of each pair (Stokes $I$ only) is inverted to provide a clearer comparison of the intensities in the FS interior.
} 
 \label{fig:ft_stokes_Ix_4lines}
\end{figure*}
\begin{figure*}
\center{{\includegraphics[width=0.95\textwidth]{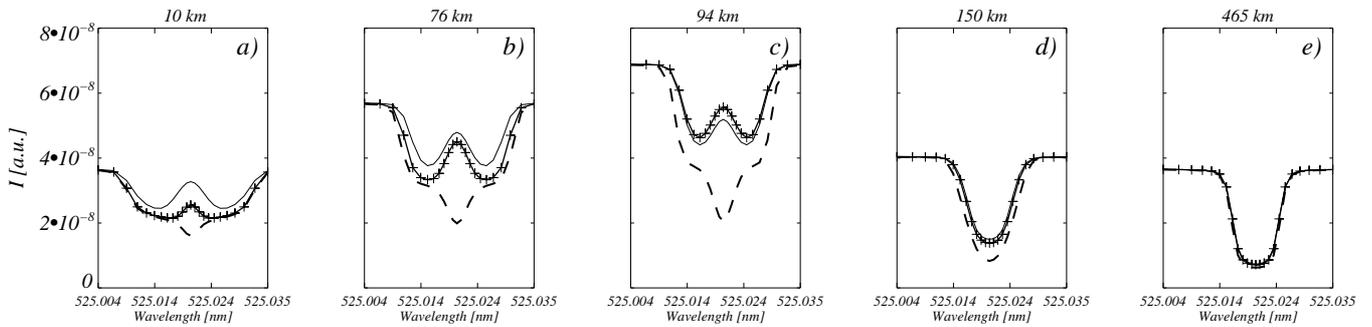}} }
 \caption{ 
 Stokes $I$ profiles of the $525.0$ nm line at five selected $x$-positions for the standard FS model including Zeeman splitting. Solid lines 3-D NLTE, dashed lines LTE, and plus signs 1-D NLTE. Refer to Fig. \ref{fig:standard_opfudge_Inu} for the profiles calculated without Zeeman splitting.
} 
 \label{fig:full_stokes_Inu}
\end{figure*}

\begin{figure*}
\center{{\includegraphics[width=0.95\textwidth]{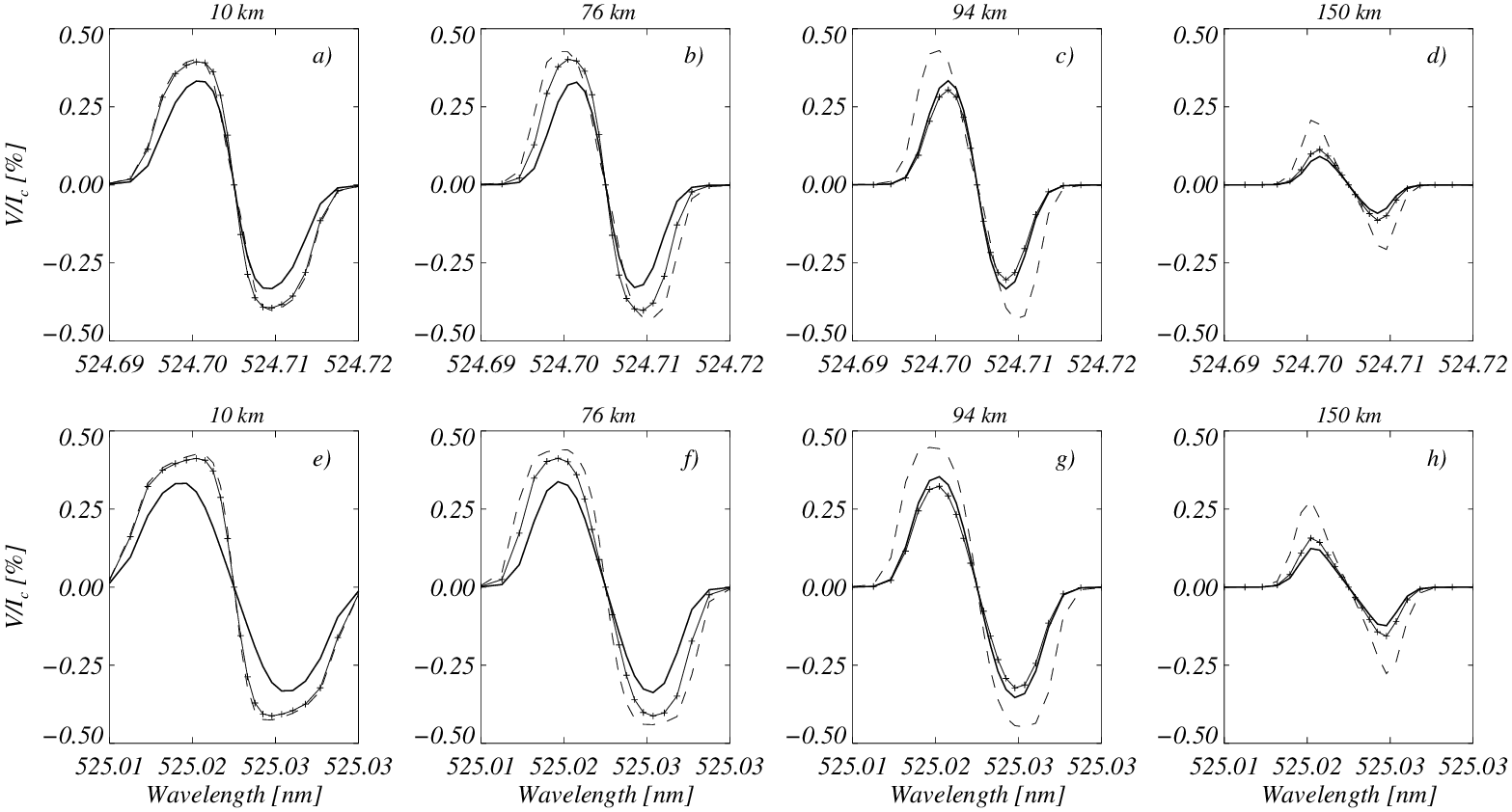}} }
 \caption{ 
 Stokes $V/I_c$ profiles of $524.7$ nm (upper panels) and $525.0$ nm (lower panels) for different $x$-positions across a FS ($I_c$ is the continuum intensity). The plotted $x$-positions correspond to the first four panels in Fig.\ref{fig:full_stokes_Inu} (Solid lines 3-D NLTE, dashed lines LTE, plus signs 1-D NLTE). 
} 
 \label{fig:full_stokes_Vnu}
\center{{\includegraphics[width=0.95\textwidth]{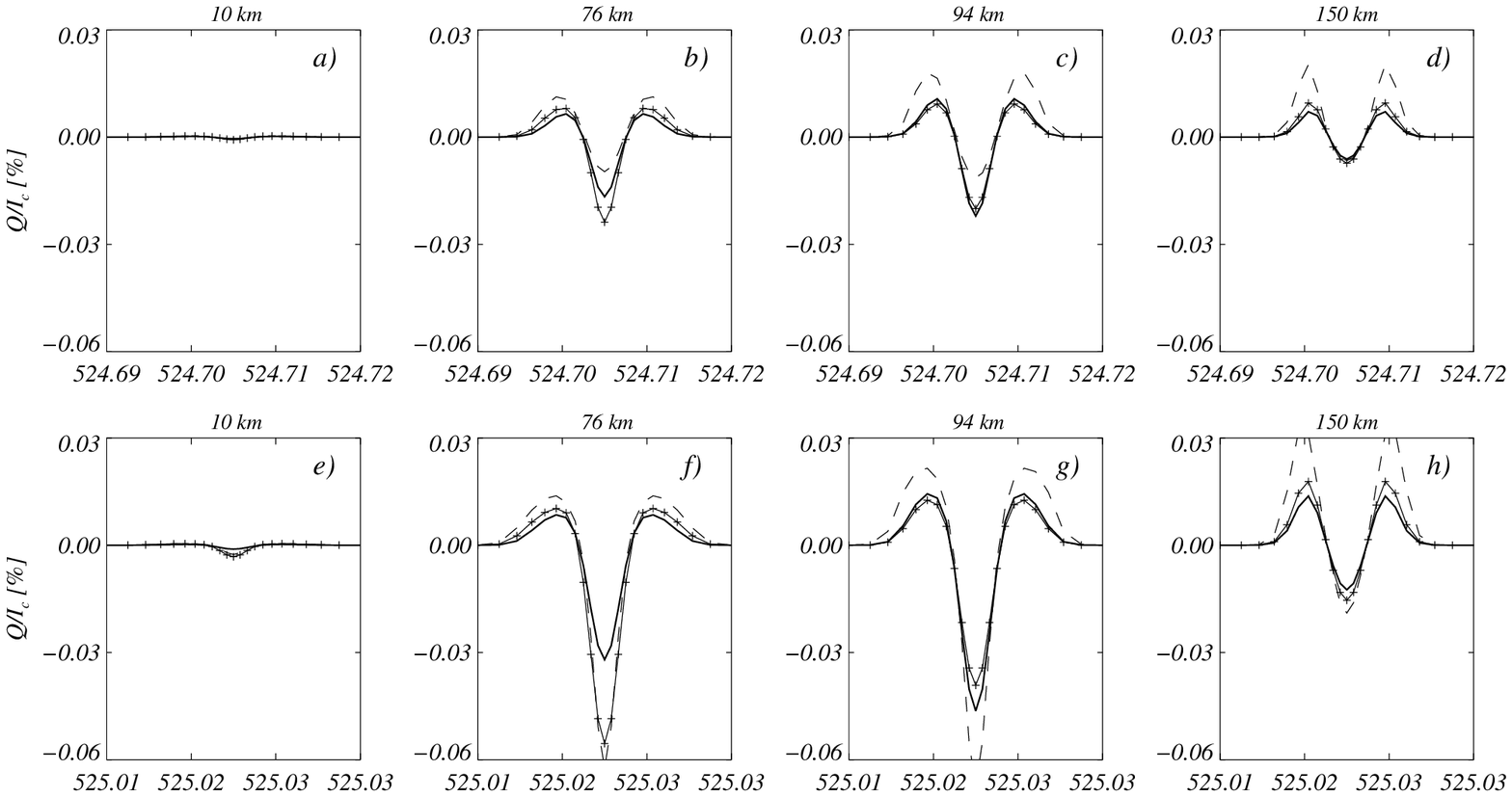}} }
 \caption{ 
 Same as Fig.\ref{fig:full_stokes_Vnu} but for Stokes $Q/I_c$.
} 
 \label{fig:full_stokes_Qnu}
\end{figure*}

Here we discuss solutions for the full Stokes vector including Zeeman splitting (following the method described in Sect. \ref{sec:fluxsheet_model_atmos} and the Appendix in more detail).

Figure \ref{fig:ft_stokes_Ix_4lines} displays the run of the line-core intensities of our four selected lines for different calculation methods.  Each panel pair compares the results obtained by including Zeeman splitting (panels on the right) with those without it (panels on the left). We note that the non-magnetic case (Stokes $I$ only) was already discussed in Sect. \ref{sec:fluxsheet_results_lines}.  We now show both the $524.7$ nm and $525.0$ nm lines as their profiles are no longer identical in the Zeeman split case due to their different Land\'e factors.

The inclusion of Zeeman splitting may lead to either an intensification or attenuation of the NLTE and horizontal RT effects, but the concrete behavior of a line can be understood in terms of its strength, Land\'e factor, and to a certain extent its temperature sensitivity.

We first consider the $524.7$ nm and $525.0$ nm lines. In the absence of Zeeman splitting these two lines are practically identical, as expected (panels a and c). In the panels b and d, however, the 
$525.0$ nm line is more strongly weakened owing to its higher Land\'e factor. This is true for all three curves.  The Zeeman splitting essentially reduces the opacity in the line core. This may allow the hot walls to become visible to the line core, as happens in LTE for the $525.0$ nm and $630.2$ nm lines, where we observe strong peaks near the boundary in the Zeeman-split case.

At the FS center, the effects of 3-D RT (i.e. the \emph{relative} difference between the 1-D and 3-D intensities) may also be either enhanced or damped. The $630.1$ nm line, which is the most Zeeman insensitive of the four considered spectral lines, displays little dependancy on Zeeman splitting in LTE and 1-D NLTE, but a strengthening of the 3-D effects. In contrast, precisely the opposite behavior is exhibited by the $630.2$ nm line, where the relative difference between the three methods is very small when Zeeman splitting is included. The $524.7$ nm and $525.0$ nm lines are particularly affected by 3-D effects. We note that these curves refer only to the line core. At other places in the line profile, even the $630.1$ nm line appears to be considerably influenced by Zeeman splitting. The exact behavior depends on details of the splitting compared to the line width, as well as other factors.

Figure \ref{fig:full_stokes_Inu} demonstrates the variation in the \ion{Fe}{i} $525.0$ nm Stokes $I$ profiles with distance from the $x$-axis of the FS, in analogy to Fig. \ref{fig:standard_opfudge_Inu}, but for the solution of the full Stokes vector. As expected, far away from the FS, the profiles computed in 3-D are very similar to LTE profiles (panel e). The commonly made assumption that \ion{Fe}{i} $525.0$ nm is "formed close to LTE" seems to be valid. The inclusion of the Zeeman effect has little influence because the line is formed mainly in the non-magnetic atmosphere below the canopy. The situation changes when we approach the boundary of the FS where the field strength is higher and the line is partially formed in the canopy. In addition to the effects already discussed in Sect. \ref{sec:fluxsheet_results_lines}, the inclusion of the full Stokes vector in the formal solution  produces much broader profiles and enhanced line-core intensities (especially in the 3-D NLTE and 1-D NLTE calculation, less so in LTE) at all locations close to or inside the FS (compare panels a) to c) of Fig. \ref{fig:full_stokes_Inu} with the corresponding panels of Fig. \ref{fig:standard_opfudge_Inu}).

Figure \ref{fig:full_stokes_Vnu} displays the Stokes $V/I_c$ and Fig. \ref{fig:full_stokes_Qnu} the $Q/I_c$ profiles corresponding to the first four $I$ profiles in Fig. \ref{fig:full_stokes_Inu}. Positions further away from the FS do not show a significant polarization signals, as the line is formed in the field-free part of the atmosphere.

The 3-D NLTE $V/I_c$ profiles are weaker than the LTE ones at all locations and, inside the FS, also weaker than the 1-D NLTE ones (except for the special case very close to the boundary, where the influence of horizontal RT is reversed). In the $525.0$ nm line, the larger Zeeman splitting in the 3-D profiles can be clearly seen (both in Stokes $I$ and $V$). In general, however, owing to 3-D RT, Stokes $V$ displays smaller effects than Stokes $I$ because these effects are largest in the line core, where Stokes $V$ is zero. Interestingly, the form of the $525.0$ nm $V$ profiles in panel e) to differ from those calculated in either 1-D or LTE or those of the $524.7$ nm line. The center of gravity of the $525.0$ nm 3-D profile shows a systematic shift away from the line center, an effect that may be of considerable relevance to the diagnostics of the magnetic field strength, since such a profile mimics a stronger field strength and a seemingly lower filling factor than the profile in LTE. The Stokes $V$ 3-D NLTE profile shape in the FS center is consistent with an enhanced weakening of the line core inside the FS.

The Stokes $Q/I_c$ profiles plotted in Fig. \ref{fig:full_stokes_Qnu}, clearly illustrate that the effect of 3-D RT is largest near the line core. At the center of the FS, these profiles vanish for reasons of symmetry (the plot does not exactly corresponds to the FS center, so that the signals do not completely disappear). Near the flux sheet's boundary, however, as shown in panel b, the strong weakening of the 3-D NLTE Stokes $Q$ in the line core is clearly apparent. In the line wings, unsurprisingly, the 1-D and 3-D profiles almost coincide. The difference in the strength of the $\pi$-components of the $524.6$/$525.0$ nm lines is due to the Zeeman desaturation of the line core, which is stronger for the higher g line.

\subsection{Variation in the field strength}\label{sec:fluxsheet_results_B}
\begin{figure*}
\center{ \resizebox{0.66\width}{0.66\height}{\includegraphics{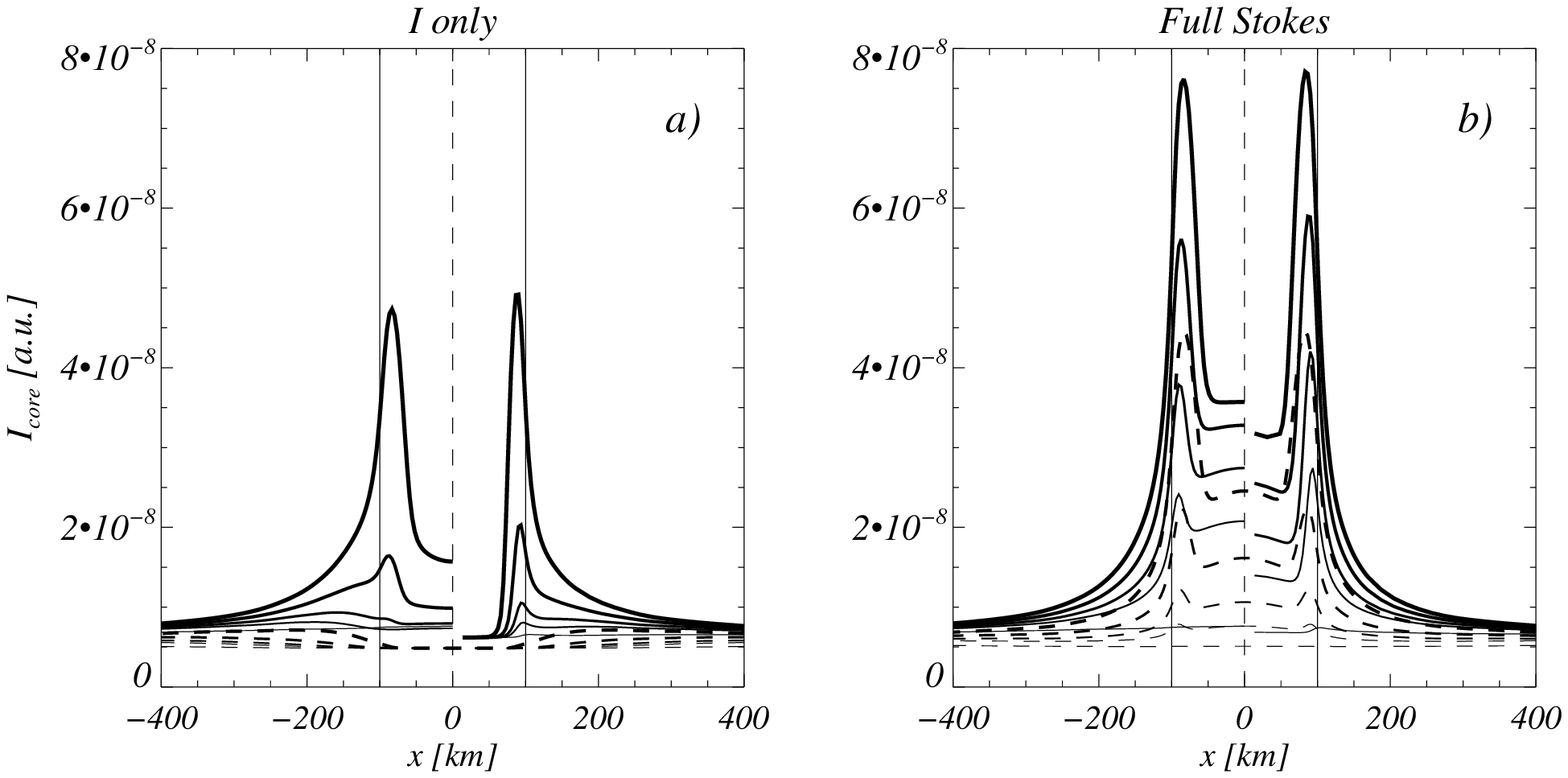}} }
 \caption{ 
 Influence of the magnetic field strength on the $525.0$ nm line-core intensity. Three-dimensional NLTE (solid, right part of each panel), 1-D (solid, left part of each panel), and LTE (dashed) for five different field strengths of 1700G (highest curves, thicker), 1600G (standard), 1500G, 1400G, and 1000G (lowest curves, thinner) at $z=0$ km. Panel a: Intensities achieved when Zeeman splitting is neglected. Panel b: those obtained after including the Zeeman effect. Vertical lines indicate the FS boundary (solid) and FS center (dashed). 
} 
 \label{fig:var_b}
\end{figure*}
\begin{table}
\caption{Wilson depression $w_D$ for different field strengths $B$ at $z=0$ km (i.e. $\tau_c=1$ in the QS) and different FT model atmospheres (NET, PLA). }
 \begin{center}
  \begin{tabular}{rrrrrr}
   \hline
   \hline
   $B$ [Gauss] \vline&  1700 &  1600  &   1500 & 1400   & 1300     \\
   $w_D$ [km]     \vline&   258 &  191   &   149  &  120   & 97        \\
   \hline
   $B$ [Gauss] \vline& 1200  & 1100   & 1000   & NET: 1600  & PLA: 1600 \\
   $w_D$ [km]     \vline& 79    & 64     & 51     & 238        & 206       \\
   \hline
  \end{tabular}
 \end{center}
\label{tab:wilson}
\end{table}

The field strength has a two-fold effect. Firstly, it affects the construction of the atmospheric models by determining the Wilson depression, which in turn influences the magnitude of the horizontal transfer effects. The Wilson depression corresponding to the different field strengths is listed in Tab. \ref{tab:wilson}. Obviously, the dependence is strongly non-linear. Secondly, the field strength directly affects the line profile through the Zeeman effect. Here we considered the effect of changing the field strength, while other atmospheric parameters were unchanged. 

Figure \ref{fig:var_b}a illustrates the influence of the magnetic field strength on the $525.0$ nm line core intensity without Zeeman splitting. As expected, the influence of the field strength is strong even when neglecting the Zeeman splitting. The main effects---the strong peak near the flux sheet's boundary and the enhanced intensity in the center of the FS---both increase with increasing Wilson depression. Below approximately $1500$ G at $z=0$ km (which corresponds to $\tau_c=1$ in the HSRA), the peak due to the hot walls disappears, while the intensity produced by 3-D RT at the FS center remains somewhat higher than in 1-D NLTE even for $1000$ G, demonstrating that horizontal radiative transport is still a relevant factor. 

Figure \ref{fig:var_b}b depicts the intensities obtained when the line splitting due to the Zeeman effect is taken  into account. The full Stokes solution provides significantly higher rest intensities close to and inside the FS compared to the solution neglecting the Zeeman effect, whereas the intensities far away are unaffected as they are mainly formed in the---non-magnetic---QS atmosphere. We note that only part of the increase in rest intensity in Fig. \ref{fig:var_b}b is directly due to Zeeman splitting. Its indirect influence in terms of 3-D RT is equally important, as can be seen by comparing Fig. \ref{fig:var_b}a with \ref{fig:var_b}b.

\section{Discussion}\label{sec:fluxsheet_discussion}

After presenting, in Sect. \ref{sec:fluxsheet_results}, the influence of various model parameters on both the line profiles and the rest intensity, we now consider questions such as, how strongly do typical observational quantities differ if one employs true 3-D RT instead of 1-D or even LTE? As \citet{stenholmstenflo1977} already pointed out, not only the inclusion of NLTE effects, but also horizontal RT may play major roles when considering a flux tube geometry. 

Here we have gone beyond the earlier studies of \citet{stenholmstenflo1977},\citet{caccinseverino1979}, and \citet{brulsvdluehe2001} by carrying out a systematic study of the changing parameters of a thin flux tube (or thin flux sheet) on the profiles of four widely used spectral lines at different levels of simplification in the RT (full 3-D NLTE, 1-D NLTE, and LTE). 

To compare the influence of different calculation methods and models on the determination of important atmospheric parameters, such as temperature, abundances, magnetic flux, and field strength, we selected frequently used observables to quantify the effect of the different parameters. We show the results for the \ion{Fe}{i} $525.0$ nm line, which is a line widely used in solar diagnostics \citep[e.g.,][]{bartholetal2011_sunrise}. The effects are similar for the other lines, though they may differ in their details. Zeeman splitting and UV opacity enhancement were included in all computations presented in this section.

\subsection{Core line depth and determination of temperature}\label{sec:discussion_linedepth}


\begin{table}
 \caption{Line depth in units of $I_c$ of the spatially averaged \ion{Fe}{i} $525.0$ nm line profile for 3-D NLTE, LTE, and 1-D NLTE computations, as well as for several model FSs derived from our standard model.  }
 \begin{center}
  \begin{tabular}[t]{lrrr}
   \hline
Model             & 3-D NLTE &\hspace{10 mm}LTE & 1-D NLTE \\
   \hline
Quiet Sun (HSRA)   &    0.83  & 0.87 & 0.83   \\  
$B=1000$ G  	   &    0.81 &   0.86 &   0.82 \\
$B=1300$ G  	   &    0.76 &   0.85 &   0.78 \\
$B=1400$ G  	   &    0.71 &   0.84 &   0.74 \\
$B=1500$ G 	   &    0.65 &   0.81 &   0.69 \\
$B=1600$ G, standard &\bf{0.59}& 0.77 &   0.63 \\
$B=1700$ G  	   &    0.53 &   0.68 &   0.56 \\
PLA ($B=1600$ G)    &    0.61 &   0.62 &   0.57 \\
NET ($B=1600$ G)     &    0.59 &   0.58 &   0.53 \\
   \hline
  \end{tabular}
 \end{center}
\label{tab:discuss_linedepth}
\end{table}

Table \ref{tab:discuss_linedepth} gives an overview over the line depth $d_l=(I_c-r)/I_c$, where $r$ is the rest intensity at line center and $I_c$ the continuum intensity, calculated for some of our model FSs. We consider here the depth of the spatially averaged profile, i.e. averaged over the area over which the flux sheet expands until it merges with its neighboring sheets. For a filling factor of 0.2, which is assumed for all computations presented in this section, this corresponds to integrating out to five times the distance from the FS boundary (at a height corresponding to $\tau=1$ in the QS). 

The results obtained for $d_l$ of \ion{Fe}{i} $525.0$ nm are summarized below and in Table \ref{tab:discuss_linedepth}. The standard model corresponds to $B=1600$ G, $R=100$ km, a filling factor of $0.2$, a wall thickness of $20$ km, the HSRA in both the QS and the FS interior, and an enhanced UV opacity. The line depth obtained in 3-D NLTE in the standard model is printed in boldface in Table \ref{tab:discuss_linedepth}.  The values for the QS entry were taken from a flat 1-D HSRA model without any magnetic structure. They correspond well to the values reached far away from the FS. We note that because of Zeeman splitting, the line core depths  do not necessarily correspond to the maximum line depth. However, in the averaged profiles, the core intensity is always very close to the minimum intensity. The main conclusions reached from these computations are: 

\begin{enumerate}
\item[1.] The average 1-D profile is always deeper than the corresponding 3-D profile except for the hotter PLA and NET atmospheres. 
\item[2.] The values of $d_l$ obtained for 1-D and 3-D RT differ by approximately $5 - 10$\%. The LTE and 3-D line depths may differ by up to 30\%. 
\item[3.] The atmosphere describing the interior of the FS is---as expected---strongly relevant. The LTE and especially the 1-D NLTE line in the hotter NET atmosphere is weaker than that of the true 3-D NLTE calculation. We note that the difference in $d_l$ between the FS atmospheres is much smaller when true 3-D NLTE is used.
\item[4.] For models with weak magnetic fields ($<1500$ G), the effects of 3-D RT are relatively small owing to the small Wilson depression and the small amount of radiation entering the FT from the walls. For strong fields, the effects increase disproportionately.
\item[5.] The influence of the variation in the geometric quantities (such as the flux sheet radius or the wall thickness, which are both not given in Tab. \ref{tab:discuss_linedepth}) on the depth of the average line profile is generally small.
\end{enumerate}

In the past, the thermal structure of the magnetic elements has generally been derived in LTE \citep[e.g.][]{stenflo1975,chapman1977,chapman1979,walton1987,solanki1986,solankibrigljevic1992}. To quantify the error in the temperature of the FS when the proper 3-D NLTE formalism was not applied, we tried to find a 1-D atmosphere that reproduces the 3-D NLTE line depth obtained for the standard FS model. As the 3-D NLTE line depth is weaker, we searched for a hotter model than the HSRA. By replacing the HSRA in the FS interior with an atmosphere whose whole temperature curve is shifted up by a constant amount relative to the HSRA, we found that the difference between the depth of the LTE and the 3-D NLTE line profiles may lead to an overestimate of the temperature of approximately $50$ K for a filling factor of 0.2.

Modern telescopes reach higher spatial resolution and can begin to resolve ever smaller magnetic elements \citep{laggetal2010}. We therefore also considered the error in the temperature obtained from observations that fully resolve the magnetic elements. If we considered the profile emerging along the center of the FS, then this temperature error could reach values as high as $300-400$ K (for the standard model). When compared with the 1-D NLTE profiles, the difference was only slightly smaller, confirming that 3-D RT is relevant as predicted by \citet{stenholmstenflo1978}. Similar temperature differences of a few 100K were also confirmed by inversions (kindly provided to us by Michiel van Noort) of the calculated spectra. As in the vast majority of the cases employed in the past, the inversions were based on LTE forward calculations. 

Although this may be an alarming result for those who determine temperatures by comparing observations with LTE calculations, one has to consider that our model is rather simple and may not correspond to a realistic scenario. Thus, we consider it important to investigate this question further using realistic 3-D radiation MHD simulations, which we plan to do in an upcoming publication. 

Interestingly, the line depth in 3-D NLTE for flux sheets whose interiors are described by the PLA and NET atmospheres is, however, rather similar to the line depths obtained in LTE for the same models. Since these models were deduced by employing the LTE approximation, this is a heartening result, since it suggests that these models may also have at least some validity for a 3-D diagnostic RT. This might help explain the similarity of the PLA model to the temperature stratification in flux sheets formed in 3-D radiation-MHD simulations \citep{vitasetal2009}, albeit the RT in the simulations is restricted to LTE.

\subsection{Equivalent width and determination of abundances}\label{sec:discussion_aeqwidth}


\begin{table}
 \caption{Same as Table \ref{tab:discuss_linedepth}, but for the equivalent width given in \% of the nominal value obtained for 3-D NLTE in our standard model (boldface font).}
 \begin{center}
  \begin{tabular}[t]{lrrr}
   \hline
Model                   & 3-D NLTE &   \hspace{10 mm}LTE   &  1-D NLTE \\
   \hline
Quiet Sun (HSRA)&        118  & 129 & 118 \\
$B=1000$ G      &        119  & 132 & 120 \\
$B=1300$ G      &        117  & 133 & 119 \\
$B=1400$ G      &        114  & 133 & 117 \\
$B=1500$ G      &        110  & 133 & 115 \\
$B=1600$ G, standard &\bf{100}& 132 & 110 \\
$B=1700$ G      &         84  & 127 &  99 \\
PLA ($B=1600$ G)	&        100  & 107 &  93 \\
NET ($B=1600$ G)	&        101  &  99 &  86 \\
   \hline
  \end{tabular}
 \end{center}
\label{tab:discuss_aeqwidth}
\end{table}
The equivalent width is a widely used observable for determining the abundances or empirical oscillator strengths. Table \ref{tab:discuss_aeqwidth} summarizes the  equivalent widths of the spatially averaged profiles for the same set of calculations and with the same averaging method as in Sect. \ref{sec:discussion_linedepth}. All the values were normalized relative to the 3-D NLTE results for the standard model, which is consequently given a value of 100\% (given in boldface). 

The typical equivalent width of the LTE lines is much larger than that of 3-D NLTE calculations (32\% in our standard model). The 1-D NLTE value is about 5 to 10\% larger. The LTE equivalent width is not strongly influenced by any variation in the model parameters because the LTE lines are typically saturated and do not vary strongly across our FS models.  Exceptions are the calculations with hotter FS/FT atmospheres (PLA, NET), where the differences between the 3-D NLTE and LTE calculation are much smaller. This is not unexpected given the results of Sect. \ref{sec:discussion_linedepth}.

Assuming that the true equivalent width is closer to that obtained in 3-D NLTE, then the larger  equivalent width obtained using LTE or 1-D NLTE can be misinterpreted as a lower iron abundance in the magnetic features or a higher temperature than really present, possibly combined with reduced microturbulence. \citet{sheminovasolanki1999} determined elemental abundances in magnetic elements under the assumption of LTE and found similar values as in the quiet Sun. However, they employed empirical model atmospheres for the magnetic elements that were derived in LTE, which likely had an incorrect temperature structure to compensate for the wrong line depths obtained in LTE. The use of these models also gives line equivalent widths close to the observed values for the correct abundances. This implies that a self-consistent LTE computations should more often lead to errors in the determined temperature than the abundances. 

By using MHD simulations with magnetic flux typical of moderate active-region plage, \citet{fabbianetal2010} found evidence of smaller equivalent widths for the iron lines in the visible, in good agreement with many previous results \citep[e.g.][]{solanki1993}. This reduced equivalent width is of relevance to the determination of abundances. However, they employed only LTE, so that they basically see the effect of the different temperature stratifications in plage and quiet regions. We expect that the use of 3-D NLTE could give rise to a sizeable additional correction.

\subsection{Contrast and feature size}\label{sec:discussion_contrast}

\begin{table}
 \caption{Same as Table \ref{tab:discuss_linedepth} but for the maximum contrast in the \ion{Fe}{i} $525.0$ nm line core (see main text for definition). Since the contrast is, by definition, zero for the quiet Sun, it has been removed.  }
 \begin{center}
  \begin{tabular}[t]{lrrr}
   \hline
Model                   & 3-D NLTE &   \hspace{10 mm}LTE   &  1-D NLTE \\
   \hline
$B=1000$ G	 	&  0.18 &   0.04 &   0.17 \\  
$B=1300$ G	 	&  1.31 &   0.19 &   1.61 \\
$B=1400$ G	 	&  2.63 &   0.49 &   3.20 \\
$B=1500$ G	 	&  4.58 &   1.26 &   5.37 \\
$B=1600$ G, standard  &\bf{7.06}&   3.04 &   7.83 \\
$B=1700$ G	 	&  9.64 &   6.80 &  10.51 \\
PLA ($B=1600$ G)		&  6.32 &   7.07 &   9.00 \\
NET ($B=1600$ G)		&  6.18 &   7.79 &   9.27 \\
   \hline
  \end{tabular}
 \end{center}
\label{tab:discuss_contrast}
\end{table}

One of the interesting results of our calculations is the strong intensity peak near the FS/FT boundary seen, e.g., in Fig. \ref{fig:standard_Ix}. We now investigate the maximum contrast, defined as $(I_{max}-I_{QS})/I_{QS}$, where $I_{QS}$ is the intensity in the quiet Sun and $I_{max}$ is the maximum intensity reached along the $x$-axis (generally near the boundary of the FS). We note that we assume an infinite spatial resolution of the observations here (in contrast to Sects. \ref{sec:discussion_linedepth} and \ref{sec:discussion_aeqwidth}). 

Although we did not find such strong contrasts as \citet{caccinseverino1979}, our results still show considerable intensity variation within the model. Table \ref{tab:discuss_contrast} gives the line-core contrast for the same model calculations as given in Sects. \ref{sec:discussion_linedepth} and \ref{sec:discussion_aeqwidth}. The contrast may be either large (strong fields, especially in  NLTE) or small (weak fields, especially for LTE), depending on whether the hot walls are in direct view of the observer. Not only do the LTE and NLTE calculations differ strongly: the 3-D and 1-D NLTE calculations may differ considerably as well as can be seen from Tab. \ref{tab:discuss_contrast}.

Another interesting finding is the very narrow geometrical width ($\approx 20$ km)  of these structures as pointed out by \citet{brulsvdluehe2001}. If such strong transitions from the FS exterior to its interior exist, a resolution of at least $5 - 10$ km is needed to see the maximum contrast. However, we stress that this test is no proof that such small structures will actually be observable, since horizontal energy transfer via radiation tends to smooth out the horizontal temperature gradients. Thus, when replacing the HSRA model in the FS by the empirically derived NET or PLA models, the large and narrow peaks in the line core intensity at the edge of the FS are either strongly damped or even disappear. This question warrants additional investigation with the help of 3-D radiation MHD simulations.

\subsection{Stokes profiles and determination of magnetic field strengths }\label{sec:discussion_stokesV}
Figures \ref{fig:full_stokes_Vnu} and \ref{fig:full_stokes_Qnu} in Sect. \ref{sec:fluxsheet_results_stokes} reveal that all Stokes parameters are strongly influenced by the calculation method and the input parameters. In Fig. \ref{fig:full_stokes_Vnu}, both the strengths of the $V/I_c$ profiles and the profile shapes differ significantly between the LTE and the 3-D NLTE case. Both quantities directly influence the determination of the magnetic field strengths. In the present section, we test this using the so-called center of gravity method \citep{reessemel1979}. The results are given in Table \ref{tab:discuss_cog} for the same models as in Table \ref{tab:discuss_contrast}.

\begin{table}
 \caption{Same as Tab. \ref{tab:discuss_contrast}, but for the field strengths $B_{cog}$, in Gauss, $10$ km from the FS center, determined by the center of gravity method in the \ion{Fe}{i} $525.0$ nm line. 
 }
 \begin{center}
  \begin{tabular}[t]{lrrr}
   \hline
Model                   & 3-D NLTE &   \hspace{10 mm}LTE   &  1-D NLTE \\
   \hline
$B=1000$ G	 	&   733 &    615 &    664 \\
$B=1300$ G	 	&  1054 &    899 &    973 \\
$B=1400$ G	 	&  1192 &   1024 &   1110 \\
$B=1500$ G	 	&  1369 &   1178 &   1280 \\
$B=1600$ G, standard 	&\bf{1627}& 1392 &   1518 \\
$B=1700$ G	 	&  2204 &   1755 &   1926 \\
PLA ($B=1600$ G)	&  1325 &   1226 &   1416 \\
NET ($B=1600$ G)	&  1349 &   1322 &   1412 \\
   \hline
  \end{tabular}
 \end{center}
\label{tab:discuss_cog}
\end{table}

For the models with the same atmosphere inside and outside the FS (HSRA/HSRA), the field strength obtained from the 3-D NLTE computations is approximately $15$\% higher than that deduced from the Stokes $V$ profiles computed in LTE, except for the $B=1700$G case, where the relative difference is larger. About half of this difference is due to horizontal RT effects, while the other half comes from NLTE (see the difference in $B_{cog}$ between LTE and 1-D NLTE in Table \ref{tab:discuss_cog}). We note that the Stokes $V$ profiles (Fig. \ref{fig:full_stokes_Vnu}) in LTE and 1-D NLTE are quite similar but that the Stokes $I$ profiles (Fig. \ref{fig:full_stokes_Inu}) differ strongly and lead to this sizeable difference in $B_{cog}$. 

Part of the difference in $B_{cog}$ between LTE and 3-D NLTE may be attributed to the different height of formation of the spectral line computed under these different assumptions. 

The use of different FS atmospheres also has a strong influence: the field strength deduced from the NET model in LTE reaches almost the value obtained from the 3-D NLTE calculation. For the slightly cooler PLA atmosphere, however, the difference in field strength between 3-D NLTE and LTE is distinctly higher, although it remains significantly smaller than that found in models with the HSRA/HSRA combination. This result further supports the conclusion that for these two empirical FT models, calculations in LTE provide a reasonable approximation for more realistic 3-D NLTE computations.

\section{Conclusions}\label{sec:fluxsheet_conclusions}

With an improved version of the 3-D RH code \citep{uitenbroek2000}, we have calculated widely used diagnostic iron lines in various flux-sheet (FS) and flux-tube (FT) models. Our adaption of the formal solution to monotonic, parabolic B\'ezier interpolation of the source function \citep{auer2003} is crucial to preventing considerable errors in the calculated intensities and to achieving better convergence. We have developed an enhanced B\'ezier interpolation scheme for the formal solution of the \emph{full} Stokes vector problem (for details see the Appendices). 

We started from a standard expanding flux-sheet model, which consisted of a HSRA atmosphere in both the non-magnetic QS and the magnetic FS, a radius $R=100$ km, field strength $B=1600$ G at external $\tau_c=1$, wall thickness $w=20$ km, and a filling factor $\alpha=0.2$. We studied the behavior of four \ion{Fe}{i} spectral lines ($524.7$ nm, $525.0$ nm, $630.1$ nm, $630.2$ nm) under three different levels of approximation for computing the radiative transfer (full 3-D NLTE RT, 1.5-D NLTE and 1.5-D LTE), each without and with the inclusion of Zeeman splitting \citep[following the \emph{field-free} method of ][]{rees1969}. We considered the dependence of these lines on FS/FT parameters, such as radius, field strength or Wilson depression, wall thickness, temperature stratification, and UV opacity.

We confirm that 3-D RT has a significant effect on spectral line formation, in qualitative agreement with the findings of \citet{stenholmstenflo1977}, who used a simple two-level atom in a cylindrical flux-tube model, and 
\citet{brulsvdluehe2001}, who employed 2-D calculations in a more realistic expanding flux-sheet geometry. In the following, we list our main conclusions.

The use of a multi-level atom seems to slightly increase the effects found by \citet{stenholmstenflo1977}. The inclusion of missing UV opacity in the way proposed by \citet{brulsetal1992}, however, strongly reduces these effects. Taken together, we found that the influence of 3-D RT is much smaller than indicated by the results of \citet{stenholmstenflo1977}.

The expanding flux sheet/tube geometry can lead to large contrasts in the line core, particularly at locations where the hot walls of the flux sheet/tube become visible \citep{caccinseverino1979}. For almost all the considered models, the NLTE computations produce larger contrasts than LTE, The contrast (and the strength of the 3-D effects) increases strongly with increasing field strength (owing to the increasing depth of the Wilson depression).

A striking feature of the emerging line radiation is the strong peak seen at the FS/FT boundary for many of the models and spectral lines, where the flux tube walls are in the direct view of the observer. These peaks (bright rings in FTs) can be a lot narrower than the generally quoted $50$ km horizontal mean-free-path for plane-parallel atmospheres even when computed in 3-D NLTE.  If such sharp boundaries exist in the solar photosphere, than they should remain visible even in the presence of horizontal RT. This result agrees with the findings of \citet{brulsvdluehe2001} and promises a new discovery space for telescopes reaching a spatial resolution as fine as $10$ km in the solar photosphere. We note, however, that one reason why the FT or FS walls were so prominently visible is that they represent a sharp horizontal transition from optically thin to optically thick gas. We suspect that the sharp structures that do not display this transition become more smeared out by the radiation.

The resulting intensity of a spherically symmetric FT model, calculated in 3-D, shows qualitatively the same effects as those of the computationally cheaper FS model. However, the influence of 3-D RT is greater in the FT than the FS owing to the larger wall area radiating into a particular point in the interior of the FT. 

Three-dimensional RT generally leads to a weakening of the line in the FS/FT interior. However, the inverse effect also occurs if the interior is hotter than the exterior (at equal geometric height), a situation occurring at certain heights if empirical models, such as the NET or PLA flux-tube atmospheres \citep{solanki1986,solankibrigljevic1992}, are used for the FS/FT interior. 

The behavior of different lines may vary strongly. The most important factors are the depth of the Wilson depression, which influences the importance of horizontal RT, and the formation height of the line, which determines whether the radiation from the hot walls reaches the absorbing atoms directly. 

The average difference between an LTE and a full 3-D NLTE RT is on the order of 10\% in the line depth and up to 30\% in equivalent width. Hence, if LTE is assumed when determining the temperature in the FS/FT interior, significant errors can result in the FT temperature when 3-D effects are not taken into account. The higher the spatial resolution, the larger the errors. For spatially resolved magnetic elements, the errors can reach $300 - 400$ K. Abundances may also be underestimated in LTE or 1-D NLTE, if correct temperatures are used for the analysis.

Non-LTE and 3-D RT effects are strong for all Stokes parameters. The difference between Stokes profiles computed under LTE, 1-D NLTE, or 3-D NLTE is of similar relative magnitude to that for Stokes $I$. The application of either LTE or 1-D NLTE is thus expected to lead to systematic errors in the determination of the magnetic field strength on the order of from 10\% to 20\% if the center of gravity method \citep{reessemel1979} is applied. Since much of this error is introduced by the line being formed at different heights in LTE and NLTE, this difference in field strength is equivalent to an error in the height to which a given field strength is assigned.

By good fortune the empirical PLA and NET models turn out to give rather similar results in LTE as in 3-D NLTE, so that the validity of results obtained with these models is broader than previously thought.

In summary, the present investigation provides evidence that the interpretation of observations of magnetic flux concentrations may change when 3-D NLTE RT effects are taken into account. Our quantitative estimates can be considered a rough guide only, because of the simplicity of the model employed here. Hence, the natural next step of this line of investigation is to employ 3-D MHD simulations and investigate commonly used diagnostics in both LTE and 3-D NLTE. Such an investigation is planned.

\begin{acknowledgements} 
We are grateful to Han Uitenbroek for providing his excellent full Stokes NLTE radiative transfer code and helpful discussions. We thank Jo Bruls for providing the iron model atom and many insights into NLTE effects. We thank Michiel van Noort for inverting some of our calculated profiles. R.~H. appreciates the flexibility of Prof. Dr. Norbert Dillier concerning the working hours at the University Hospital of Zurich. This work has been partially supported by WCU grant No. R31-10016 funded by the Korean Ministry of Education, Science and Technology. 
\end{acknowledgements}

\bibliographystyle{aa}
\bibliography{journals,holzreuter}

\begin{appendix} 
\newpage
\newpage

\section{Formal solution: Influence of the source function interpolation method}\label{sec:formal_rte_Method}

In the short characteristics (SC) method \citep{olsonkunasz1987,kunaszauer1988}, the formal solution is integrated only along the "short" path from the nearest up- and downwind points. The incoming intensities---as well as all other quantities used for the formal integration at the up- and downwind points---have usually to be interpolated from neighboring grid points. The problem of polynomial interpolation at the up- and downwind points, which may lead to spurious extrema, was widely discussed in the literature until \citet{auerpaletou1994} found an elegant and efficient solution using monotonic interpolation.

In contrast, one finds very few references to the problems that arise when the source function is interpolated with a parabolic Lagrange polynomial along the short characteristic as proposed and established as standard by \citet{aueretal1994}. With $S(\tau)=S_0+c_1\tau+c_2\tau^2$, the integral over $S$ along the short characteristics is then given by the analytical expression
\begin{equation}
   I_{local} = c_0S_0+c_1w_1+c_2w_2
\rm \ ,
\end{equation}
where $I_{local}$ is the locally produced intensity without irradiation from outside. The $w_i$ may be---in the case of Lagrange interpolation--- expressed as simple functions of known quantities \citep[see ][]{auerpaletou1994}.
The first indication of these interpolation problems is given by \citet{auer2003}, who proposes monotonic B\'ezier interpolation instead of simple Lagrange interpolation to prevent spurious extrema of S along the integration path. 

\begin{figure}
\center{\includegraphics[width=0.35\textwidth]{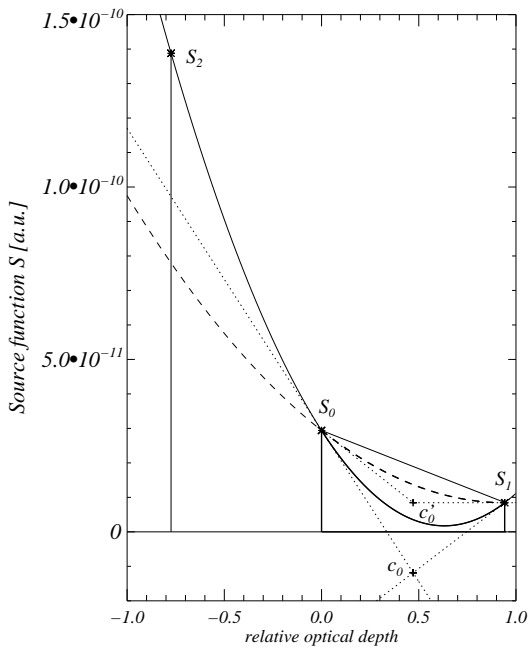}}
 \caption{ 
 Interpolation methods: The Lagrange parabola (solid) does not give a good representation of the run of S along the short characteristic from $S_0$ to $S_1$. Parabolic monotonic B\'ezier interpolation (dashed) is achieved by 
restricting the control point $c_0$ (defined by the two tangents (dotted) of each parabola) to $S_0 <= c_0^{'} <= S_1$.
} 
 \label{fig:bez_vs_lagr}
\end{figure}

To demonstrate that the proper interpolation of $S$ is as important as the correct interpolation of up- and downwind quantities, we present the situation of an overshooting source function, which we call "a critical case", taken from our fluxsheet atmosphere simulation (the standard model, see Sect. \ref{sec:fluxsheet_model_atmos} for a description). Fig. \ref{fig:bez_vs_lagr} shows the values of $S$ at the up- and downwind points ($S_1, S_2$) and the integration point ($S_0$). The unique Lagrange parabola through the three base points is given by the solid line. One immediately sees that $S$ clearly takes values below the base points $S_0$ and $S_1$, while it should decrease monotonically from $S_0$ to $S_1$. From 
\begin{equation}\label{eq:formal}
\vec{I}
\,\,=\,\,
\vec{I}_{1}
e^{-\Delta \tau_1}\,+\,
\int\limits_{0}^{\tau_1}  S(\tau)e^{-\tau}d\tau
\rm \ ,
\end{equation}
it directly follows that the resulting intensity must be wrong (i.e. too small), because the area under $S$ is far too small. We note that the factor $e^{-\tau}$ does not change the qualitative result as it is on the order of unity over the relevant integration interval. 

B\'ezier parabolas are defined by two base point values and a single control point,which is always located at the intersection of the two tangents of $S$ at the base points. This control point therefore "controls" the run of S between the two base points. The unique Lagrange parabola through $S_0$, $S_1$, and $S_2$ is recovered in the interval from $S_0$ to $S_1$ when the control point ($c_0$ in Fig. \ref{fig:bez_vs_lagr}) lies on the corresponding tangents of the Lagrange parabola (fine dotted). One imposes monotonicity in the interval from $S_0$ to $S_1$ by constraining the control point to values between the two base point values of $S$. By moving the control point up to the lower of the two values ($c_0^{'}$), one gets a new parabola (that no longer passes through $S_2$!) with a monotonic run in the integration interval (dashed curve). Therefore, the area under $S$ is no longer artificially reduced. The difference in the locally produced intensity for our sample, taken from an existing calculation, is considerable. 

We therefore used monotonic parabolic B\'ezier interpolation as proposed by \citet{auer2003} for all our calculations. We performed tests with a snapshot taken from a realistic 3-D radiation hydrodynamic simulation run (i.e. for $B=0$) with the MURAM code \citep{voegleretal2005}. Additionally, we used a 3-D version of the FAL-C model of \citet{fontenlaetal1993} obtained by arranging many identical 1-D atmospheres into a 3-D cube. In both cases, we found that about 3\% of all formal solutions suffer from $S$ overshooting in the integration interval. While the resulting intensity for the FAL-C model is almost the same as for the Lagrange interpolation (most of the critical cases in FAL-C display only quite weak overshooting because of the smoothness of the atmosphere), the resulting intensities in the snapshot may differ considerably owing to the strongly varying atmospheric parameters along the integration path. Therefore, the source function may exhibit strong overshooting.

An increasing number of publications have recognized this problem in recent years. For instance, \citet{HauschildtBaron2006} used linear interpolation (instead of monotonic B\'ezier) for critical cases and \citet{Koesterkeetal2008}, and \citet{Hayek2008} used cubic B\'ezier polynomials.

In more extreme atmospheres than our fluxsheet model (as found e.g. in realistic MHD simulations, which will be discussed in future publications), the resulting intensity may even be wrong by orders of magnitude with disastrous consequences not only for the final result, but also for the convergence of the RT computations themselves. The convergence is affected because the (usually diagonal) approximate Lambda operator used for the improvement of the iteration is generally interpolated in the same manner as the source function.

In our experience, we have found that the monotonic, parabolic B\'ezier interpolation, which is identical to Lagrange interpolation when S does not overshoot in the integration interval, has much better convergence properties than the pure Lagrange interpolation, which can lead to a complete divergence for non-smooth atmospheres, such as those along rays passing through the walls of a flux tube. 

However, imposing either monotonicity or a special treatment in critical cases  may lead to convergence problems as well because a change in the interpolation method at a certain point in the cube may influence neighboring points in such a way, that in the next iteration the interpolation method is changed back again to the original method. This may lead to an infinite flip-flop situation, so that convergence is never achieved. We encountered these flip-flop situations in a few of our unsmoothed flux-sheet models. They occurred at a convergence level of a few percent in the population numbers and were much stronger when linear interpolation was used in the critical cases (a change from linear to parabolic may have more influence on the area under $S$ than a change from monotonic B\'ezier to normal Lagrange parabolas).

The  problem of strongly overshooting source functions may also arise in the 1-D schemes as they interpolate $S$ in the integration interval with a Lagrange polynomial as well. In addition, we therefore implemented and used monotonic B\'ezier interpolation for our 1-D calculations.

\section{Formal solution for the full Stokes vector problem}\label{sec:formal_rte_fullStokes}
\begin{figure}
\center{\includegraphics[width=0.45\textwidth]{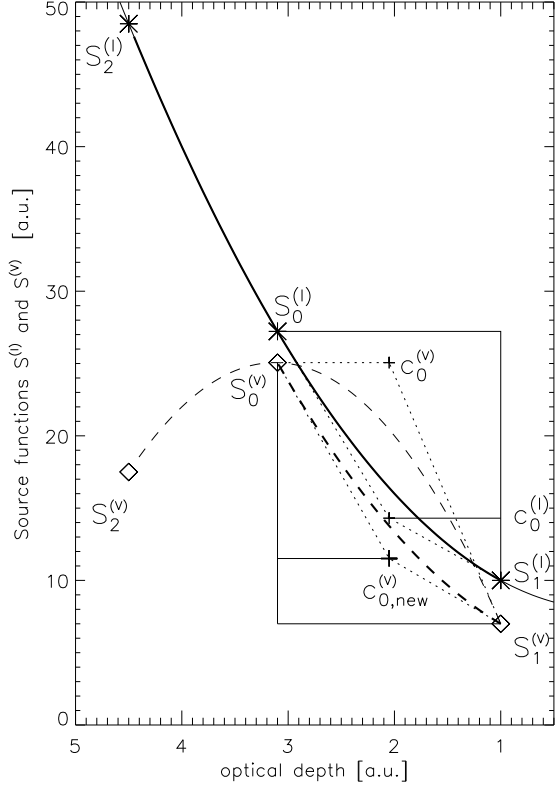}}
 \caption{  
 Interpolation method for the full Stokes vector problem. Even with monotonic B\'ezier interpolation, the source function of a Stokes component, e.g. $S_V$ (dashed thin) may become larger than $S_I$ (solid thick) in the range from $S_0$ to $S_1$. The standard control point $c_0^{(V)}$ lies far above $c_0^{(I)}$ (thin plus signs, both defined by two tangents given by thin dotted lines) owing to the different curvatures of the source function parabolas. To prevent these situations, we chose the control point for the Stokes $V$ component ($c_{0,new}^{(V)}$, thick plus sign), according to $c_0^{(I)}$. Our $c_{0,new}^{(V)}$ is defined by the ratio $(S_0^{(x)}-c_0^{(x)}) / (S_1^{(x)}-c_0^{(x)})$ (with $x$ equal to $I$ or $V$), which has to be constant. This choice prevents the source function integrals of the Stokes components from becoming larger than those of Stokes $I$.
} 
 \label{fig:bez_stokes}
\end{figure}

The solution to the formal integration of the RT equation for the \emph{full} Stokes vector within the short characteristics method of \citet{olsonkunasz1987} and \citet{kunaszauer1988} was given by \citet{reesetal1989}. Their second method, which is based on a diagonal-element lambda operator (DELO) uses linear interpolation of the source function components in the integration interval. \citet{socasnavarroetal2000} improved the accuracy of the source function integration by introducing parabolic interpolation. The RH code is based on this improved method.

As mentioned in the previous section, the interpolation with a Lagrange parabola may cause great difficulty in the formal solution. However, these problems are not limited only to Stokes $I$. The overshooting of the parabola may affect the other Stokes components in the same way. We therefore used the monotonic B\'ezier interpolation for all components.

Unfortunately, the full Stokes vector problem is even more intricate: the integration with a monotonic, B\'ezier-interpolated source function may still lead to unphysical results in special (but not uncommon) cases. Fig. \ref{fig:bez_stokes} illustrates a pathological situation where the integration of the source functions  leads to the unphysical result of $V>I$. Owing to the different (in this case even opposite) curvatures of $S_I$ and $S_V$ (the latter being a representative of all Stokes components), $S_V$ is larger than $S_I$ for a large part of the integration interval ($\tau_0$ to $\tau_1$). The likelihood of this occurring is highest when $S_V$ is almost as large as $S_I$ at one point and much smaller (or negative) at neighboring point(s). 

We note that the problem arises because one always has to consider a certain neighborhood of the integration interval when determining the interpolating function (except for linear interpolation where the two end points are sufficient). This very general result of interpolation theory is also responsible for the overshooting of $S_I$ in the case of parabolic Lagrange interpolation. If the neighborhood does not behave "well", it may produce unrealistic behavior (i.e. oscillations or overshooting) in the integration interval.

The B\'ezier parabola helps to reduce the influence of these difficult neighborhoods as it is defined in a more local way, i.e. by the two end points and a local control point, thus limiting the influence of points outside the integration interval.  We note that the control point is influenced by the neighborhood (i.e. $S_2$), but in a less strict way. By imposing monotonicity (i.e. the control point has to be between the two values of the endpoints), the influence of the neighborhood is reduced exactly for these cases where overshooting would otherwise take place.

As all Stokes components contribute to  Stokes $I$, it is reasonable to take the shape of $S_I(\tau)$ as a reference. If we replace $c_0^{(V)}$ by $c_{0,new}^{(V)}$ in such a way that the shape of $S_I(\tau)$ predefines the shape of $S_V(\tau)$ along the integration interval, the unphysical condition of $S_V>S_I$ may be prevented. For B\'ezier parabolas, this can easily be achieved with
\begin{equation}
   \frac{S_0^{(QUV)} - c_{0,new}^{(QUV)}}{c_{0,new}^{(QUV)} - S_1^{(QUV)} }
  =\frac{S_0^{(I)}   - c_0^{(I)}        }{c_0^{(I)}         - S_1^{(I)}   } 
\rm \ .
\end{equation}
In this way, we use only local information, i.e. the shape of $S_I(\tau)$ and the endpoint values of $S_{QUV}$, to define the run of $S_{QUV}(\tau)$. 

We note that in the DELO method of \citet{reesetal1989}, the integration of the source functions is only part of the solution. The  determination of the full Stokes vector intensities requires the solution of a system of linear equations. Even with our integration method, in very rare cases, the solution of the equation system produces unphysical results with one or more Stokes components becoming larger than Stokes $I$. Consequently, we changed the RH code in such a way, that the polarized Stokes component fulfills the condition
\begin{equation}
   \sqrt{ Q^2 + U^2 + V^2 } \leq I 
\rm \ 
\end{equation}
after each formal solution. In all of these cases in which the relation was not fulfilled, the polarized components were reduced by a single factor $f$ (i.e. $Q_{new}=fQ$, $U_{new}=fU$, $V_{new}=fV$) such that the equality in the equation above held, i.e. 
\begin{equation}
    \sqrt{ {Q_{new}}^2 + {U_{new}}^2 + {V_{new}}^2 } = I 
\rm \ .
\end{equation}

\end{appendix}

\end{document}